\documentclass{article}
\usepackage{jheppub}
\usepackage{lineno}
\usepackage{mathtools}
\usepackage{booktabs}
\usepackage{needspace}
\usepackage{placeins}
\usepackage{tikz}
\title{\boldmath A Dispersive Bootstrap for the Virasoro-Shapiro Amplitude}

\author[a,b,c]{Yongjun Xu}

\affiliation[a]{School of Fundamental Physics and Mathematical Sciences,\\
Hangzhou Institute for Advanced Study, UCAS, Hangzhou 310024, China}

\affiliation[b]{Institute of Theoretical Physics, Chinese Academy of Sciences,\\
Beijing 100190, China}

\affiliation[c]{University of Chinese Academy of Sciences,\\
Beijing 100049, China}

\emailAdd{xuyongjun23@mails.ucas.ac.cn}

\abstract{We study the closed-string tree-level Virasoro-Shapiro amplitude using the
dispersive S-matrix bootstrap. For the ten-dimensional maximally
supersymmetric four-point amplitude, we impose analyticity, crossing symmetry,
partial-wave unitarity, and Regge boundedness. With the massless graviton pole
kept explicitly, the resulting dispersion relations and crossing null
constraints give numerical bounds on the leading low-energy coefficients
normalized by the gravitational coupling, which agree at our reference
truncation with earlier bounds from smeared functionals. We then introduce a
Virasoro-inspired ansatz, which becomes a set of nonlinear
relations among Wilson coefficients and shrinks the allowed region toward the
Virasoro-Shapiro trajectory. Finally, we study a gravity-pole-subtracted
setup, where the regular part of the amplitude has a well-defined forward
limit. In this stripped problem, on the slice
of the leading coefficient that places the first Virasoro-Shapiro
resonance at the cutoff, the nonlinear constraints reduce the allowed
region to a small island containing the Virasoro-Shapiro point. Imposing the
ansatz to all orders, together with spectral assumptions not used in the
numerics, we show analytically that positivity selects the Virasoro-Shapiro
spectrum.
}


\newcommand{\dd}{\mathrm{d}}
\newcommand{\M}{\mathcal{M}}

\newcommand{\K}{\mathcal{K}}
\newcommand{\avg}[1]{\left\langle #1\right\rangle}

\begin{document}
\maketitle
\flushbottom

\section{Introduction and setup}
\label{sec:intro-setup}

The modern S-matrix bootstrap revisits the old bootstrap philosophy in a
sharper and more quantitative form. Basic principles of scattering amplitudes,
such as analyticity, crossing symmetry, unitarity, and Regge boundedness, place
nontrivial constraints on low-energy effective field theories
\cite{Adams:2006sv,Tolley:2020gtv,Caron-Huot:2020cmc,Paulos:2017fhb,Sinha:2020win,Guerrieri:2021tak}.
A particularly useful way to implement these principles is through dispersive
sum rules
\cite{deRham:2025vaq,Pham:1985cr,Pennington:1994kc,Nicolis:2009qm,Komargodski:2011vj,Remmen:2019cyz,Bellazzini:2019xts,Herrero-Valea:2019hde,Bellazzini:2020cot,Bellazzini:2017fep,Alberte:2020jsk,deRham:2017imi,Wang:2020jxr,Alberte:2020bdz,Tokuda:2020mlf,Li:2021lpe,Caron-Huot:2021rmr,Du:2021byy,Bern:2021ppb,Li:2022rag,Caron-Huot:2022ugt,Saraswat:2016eaz,Arkani-Hamed:2021ajd,Herrero-Valea:2020wxz,Guerrieri:2021ivu,Henriksson:2021ymi,EliasMiro:2022xaa,Bellazzini:2021oaj,Herrero-Valea:2022lfd,Hong:2023zgm,Chiang:2022jep,Huang:2020nqy,Noumi:2021uuv,Xu:2023lpq,Chen:2023bhu,Noumi:2022wwf,deRham:2022hpx,Hong:2024fbl,Bern:2022yes,Ma:2023vgc,DeAngelis:2023bmd,Acanfora:2023axz,Aoki:2023khq,Xu:2024iao,EliasMiro:2023fqi,McPeak:2023wmq,Riembau:2022yse,Caron-Huot:2024tsk,Caron-Huot:2024lbf,Wan:2024eto,Berman:2024owc,Beadle:2024hqg,Bellazzini:2025shd,Ye:2025zhs,Bonnefoy:2025uzf,Bellazzini:2025bay,Gumus:2025hwq,Fernandez:2026fby}.
When combined with partial-wave positivity, these sum rules express
low-energy Wilson coefficients as positive moments of high-energy spectral
data.

String amplitudes provide a natural testing ground for this rigidity. The color-ordered open-string four-point amplitude is governed by the Veneziano amplitude \cite{Veneziano:1968yb}. Recent work has studied to what extent this amplitude is fixed by general bootstrap principles supplemented by mild stringy input. In particular, positivity bounds combined with string monodromy have been shown to carve out small islands around the open-string point \cite{Huang:2020nqy}. Positivity combined with higher-point splitting relations and hidden-zero conditions has also been shown to shrink the allowed region toward the Veneziano amplitude \cite{Berman:2025owb}. These stringy inputs were later shown
to be sufficient, within an analytic moment-problem bootstrap, to fix the
Veneziano amplitude uniquely \cite{Wan:2026pjq}. Other approaches study
deformations of the Veneziano amplitude, level truncation, improved Regge
softness, and the emergence of the string spectrum from bootstrap assumptions
\cite{Geiser:2022exp,Cheung:2022mkw,Cheung:2023adk,Cheung:2024uhn}. Taken
together, these results suggest that the Veneziano amplitude is not an
arbitrary solution, but rather a highly rigid point in the space of consistent
tree-level amplitudes.

The analogous question for closed strings is more subtle because gravity is already present at tree level. The Virasoro-Shapiro amplitude contains the massless graviton pole, which obstructs the standard forward-limit positivity argument. One way to deal with this obstruction is to replace the strictly forward observable by a smeared impact-parameter observable, or equivalently by scattering suitable wave packets \cite{Caron-Huot:2021rmr,Caron-Huot:2022ugt}. Another approach is to keep the gravity pole explicitly in the dispersion relation and sample it directly together with the discretized spectral density \cite{Peng:2026ztp}.
In this work we consider the simpler tree-level setup of maximally supersymmetric ten-dimensional gravity, studied in \cite{Caron-Huot:2021rmr} and analysed systematically in \cite{Albert:2024yap}. In this framework, Type II string theory lies inside the allowed region, while some extremal solutions exhibit approximately string-like Regge trajectories \cite{Albert:2024yap}. Related analytic work has studied deformations of the Virasoro-Shapiro amplitude and possible uniqueness criteria for it \cite{Geiser:2022exp,Cheung:2024obl,Cheung:2025tbr}.

In this study we ask how much of the closed-string Virasoro-Shapiro amplitude
can be recovered from a tree-level S-matrix bootstrap. We focus on the
ten-dimensional maximally supersymmetric four-point amplitude. First, we keep
the massless gravitational pole explicitly and impose fixed-\(t\) dispersive
sum rules, partial-wave unitarity, Regge behavior, and crossing null
constraints \cite{Caron-Huot:2020cmc,Caron-Huot:2021rmr,Peng:2026ztp}. The
leading bounds at our reference numerical settings agree with those of
ref.~\cite{Albert:2024yap}, and we compare them with the exact
Virasoro-Shapiro trajectory. The new ingredient of this work is a
Virasoro-inspired triple-crossing ansatz. In the low-energy expansion, this
ansatz becomes a set of algebraic nonlinear relations among Wilson
coefficients, and we study how these relations shrink the allowed region.

Finally, we consider a gravity-pole-subtracted setup. After removing the
massless pole, the regular part of the amplitude has a well-defined forward
limit, and the moment-problem structure becomes more transparent. In this
stripped problem, at \(c_{0,0}M^6/(8\pi G)=2\zeta(3)\), the nonlinear
relations reduce the allowed region to a small island containing the
Virasoro-Shapiro point. This slice places the first Virasoro-Shapiro
resonance at the cutoff. Smaller values of \(c_{0,0}M^6/(8\pi G)\) are
equivalent to raising the spectrum above the cutoff and give larger regions
(section~\ref{subsec:stripped-numerics}). We then analyse the ansatz imposed to all orders, which is a stronger
problem than the one solved numerically. Partial-wave positivity at nonzero
momentum transfer forces every isolated exchange to enter with unit weight.
With two further assumptions that are not imposed numerically, the pole
locations are equally spaced and the coefficients take their Virasoro-Shapiro
values. This does not show that the numerical island shrinks to a point, but
it identifies the mechanism behind it: the ansatz predicts non-forward data
from forward data, which positivity then tests. The assumptions and results
are summarized in table~\ref{tab:assumptions-results}.
\begin{table}[htbp]
\centering
\begin{tabular}{p{0.23\textwidth}p{0.69\textwidth}}
\toprule
Result & Physical assumptions \\
\midrule
Dispersive bounds with gravity & Analyticity, crossing symmetry, partial-wave
unitarity, Regge boundedness, and a mass gap. \\
NL3 and NL6 bounds & The preceding physical assumptions and the
Virasoro-inspired coefficient relations:
\eqref{eq:ansatz-NL1}--\eqref{eq:ansatz-NL3} for NL3, or
\eqref{eq:ansatz-NL1}--\eqref{eq:ansatz-NL6} for NL6. \\
Pole-subtracted region & The preceding dispersive assumptions, a regular
forward limit with positive moments, the NL3/NL6 relations, and
\(c_{0,0}M^6/(8\pi G)=2\zeta(3)\). \\
Analytic string spectrum & The all-order ansatz
\eqref{eq:virasoro-inspired-ansatz}, positive forward moments and
non-forward partial waves, isolated massive exchanges with no finite mass
accumulation, and finitely many spins at the first massive level. \\
\bottomrule
\end{tabular}
\caption{Physical assumptions in the ten-dimensional maximally supersymmetric
tree-level setup. The additional spectral hypotheses in the last row are
used only in the all-order analytic argument.}
\label{tab:assumptions-results}
\end{table}

We now specify the amplitude and the bootstrap assumptions. Maximal
supersymmetry organizes the four-point amplitude into a superamplitude of the
schematic form
\begin{equation}
  \mathcal{A}_4
  =
  \delta^{10}(P)\,\delta^{16}(Q)\,\M(s,t,u),
  \qquad
  s+t+u=0 .
\end{equation}
Here \(\delta^{10}(P)\) enforces momentum conservation, while
\(\delta^{16}(Q)\) is the supermomentum-conserving delta function for the
maximal ten-dimensional massless multiplet. All external states, including the
graviton, dilaton, antisymmetric tensor, and their superpartners, belong to the
same on-shell supermultiplet. Supersymmetry Ward identities fix the Grassmann
and tensor structure of the four-point amplitude, leaving a single scalar
function \(\M(s,t,u)\) as the dynamical object to be bootstrapped. For further
details on this setup, see \cite{Albert:2024yap}.

The scalar function \(\M(s,t,u)\) depends only on the Mandelstam invariants and
is symmetric under permutations of \(s,t,u\). This is the main simplification
provided by maximal supersymmetry: instead of bootstrapping many independent
polarization structures, the problem reduces to a bootstrap for one crossing
symmetric scalar function with a positive partial-wave expansion.

At tree level, analyticity means that, at fixed momentum transfer, the only
singularities of \(\M\) are real poles or cuts associated with physical
intermediate states. Unitarity is imposed through a positive partial-wave
decomposition of the discontinuity,
\begin{equation}
  {\rm Im}\,\M(s,t)
  =
  s^{(4-D)/2}
  \sum_{\ell=0,2,\ldots}
  n^{(D)}_\ell\,
  \rho_\ell(s)\,
  P_\ell^{(D)}
  \left(1+\frac{2t}{s}\right),
  \qquad
  \rho_\ell(s)\ge 0 .
  \label{eq:partial-wave}
\end{equation}
Here \(P_\ell^{(D)}\) is the Gegenbauer-normalized polynomial
\begin{equation}
  P_\ell^{(D)}(x)
  =
  {}_2F_1\left(
  -\ell,\ell+D-3;\frac{D-2}{2};\frac{1-x}{2}
  \right),
  \qquad
  P_\ell^{(D)}(1)=1 ,
\end{equation}
and the positive normalization factor is
\begin{equation}
  n_\ell^{(D)}
  =
  \frac{
  (4\pi)^{D/2}
  (D+2\ell-3)\,
  \Gamma(D+\ell-3)}
  {
  \pi\,
  \Gamma\!\left(\frac{D-2}{2}\right)
  \Gamma(\ell+1)}
  .
  \label{eq:partial-wave-normalization}
\end{equation}
Throughout the note we use the bracket notation
\begin{equation}
  \avg{\cdots}
  \equiv
  \sum_{\ell=0,2,\ldots}^{\infty}
  n_\ell^{(D)}
  \int_{M^2}^{\infty}
  \frac{\dd s'}{\pi s'}\,
  (s')^{2-D/2}\,
  \rho_\ell(s')\,
  (\cdots) .
  \label{eq:bracket-definition}
\end{equation}

For massless external states,
\begin{equation}
  s+t+u=0 ,
\end{equation}
and it is convenient to introduce the crossing-invariant variables
\begin{equation}
  x\equiv st+su+tu
  =
  -\frac{1}{2}\left(s^2+t^2+u^2\right),
  \qquad
  y\equiv stu,
  \qquad
  a\equiv \frac{y}{x}.
  \label{eq:crossing-invariants}
\end{equation}
The amplitude may therefore be regarded as a function \(\M(x,a)\). The mass
gap \(M\) defines the cutoff scale below which all massive  states are
integrated out. In this low-energy regime the amplitude is described by maximal
supergravity plus local higher-derivative corrections. The existence of the gap
implies that the only singularity of \(\M(s,t,u)\) in a neighbourhood of the
origin is the massless graviton pole; the remaining part is analytic and admits
a crossing-symmetric Taylor expansion for
\(\max(|s|,|t|,|u|)<M^2\). Thus we write
\begin{equation}
  \M_{\rm low}(x,a)
  =
  \frac{8\pi G}{a x}
  +
  \sum_{m=0}^{\infty}\sum_{n=0}^{m} c_{m,n} x^m a^n .
  \label{eq:low-expansion-xa}
\end{equation}
The pole at \(a=0\), equivalently \(stu=0\), comes from long-range graviton
exchange. The remaining terms are local contact interactions in the effective
field theory. Locality requires \(n\le m\), since
\(x^m a^n=x^{m-n}y^n\) must be polynomial in the crossing variables \(x\) and
\(y\). Keeping the first few terms gives
\begin{equation}
  \M_{\rm low}(s,t,u)
  =
  \frac{8\pi G}{stu}
  +c_{0,0}
  +c_{1,0}x
  +c_{1,1}xa
  +c_{2,0}x^2
  +c_{2,1}x^2a
  +c_{2,2}x^2a^2
  +\cdots .
  \label{eq:low-expansion-first-terms}
\end{equation}

\section{Sum rules, crossing, and numerical setup}
\label{sec:sum-rules}

\subsection{Fixed-\texorpdfstring{\(t\)}{t} dispersion relations}
\label{subsec:fixed-t}

For fixed \(t<0\), we assume the improved Regge behavior of the
supersymmetric scalar amplitude. This allows us to consider the contour
integrals
\begin{equation}
  \frac{1}{2\pi i}
  \oint_\infty
  \frac{\dd s}{s}\,
  \frac{\M(s,t)}{[s(s+t)]^{k/2}}
  =0,
  \qquad
  k=-2,0,2,4,\ldots .
  \label{eq:fixed-t-contour}
\end{equation}
The large contour in the complex \(s\)-plane can be deformed onto the
\(s\)- and \(u\)-channel cuts, together with small contours around the
low-energy graviton poles at \(s=0\) and \(s=-t\). The Regge assumption removes
the contribution from infinity. The small contours give the low-energy Taylor
coefficients or pole residues, while the two cuts are related by crossing and
combine into a positive UV average. The resulting fixed-\(t\) kernel is
\begin{equation}
  \K_k(\mu,t,\ell)
  =
  \frac{2\mu+t}{\mu+t}\,
  \frac{1}{[\mu(\mu+t)]^{k/2}}\,
  P_\ell^{(D)}\left(1+\frac{2t}{\mu}\right).
  \label{eq:fixed-t-kernel}
\end{equation}

The first rows of the fixed-\(t\) dispersion relation are
\begin{align}
  k=-2:\qquad
  -\frac{8\pi G}{t}
  &=
  \avg{\K_{-2}}
  =
  \avg{
  \mu(2\mu+t)\,
  P_\ell^{(D)}\left(1+\frac{2t}{\mu}\right)
  },
  \label{eq:kminus2-row}
  \\
  k=0:\qquad
  c_{0,0}
  -c_{1,0}t^2
  +c_{2,0}t^4
  +\cdots
  &=
  \avg{\K_0}
  =
  \avg{
  \frac{2\mu+t}{\mu+t}\,
  P_\ell^{(D)}\left(1+\frac{2t}{\mu}\right)
  },
  \label{eq:k0-row}
  \\
  k=2:\qquad
  -c_{1,0}
  -c_{1,1}t
  +2c_{2,0}t^2
  +\cdots
  &=
  \avg{\K_2}
  =
  \avg{
  \frac{2\mu+t}{\mu(\mu+t)^2}\,
  P_\ell^{(D)}\left(1+\frac{2t}{\mu}\right)
  } .
  \label{eq:k2-row}
\end{align}

For \(k\ge0\), we do not impose the full \(t\)-dependent dispersion relations
by sampling in \(t\). Instead, we expand the kernels around \(t=0\). This
expresses the EFT Wilson coefficients as UV moments. Whenever the same Wilson
coefficient is obtained from two different Taylor coefficients or from two
different dispersion rows, the difference between the corresponding UV moment
representations must vanish. These consistency conditions are the crossing null
constraints. They were constructed systematically in
ref.~\cite{Caron-Huot:2020cmc}. Ref.~\cite{Caron-Huot:2021rmr} then
introduced improved sum rules, which add null constraints to the ordinary sum
rules. Without gravity they contain the same information as the ordinary sum
rules and null constraints together. Their advantage is that they remain well
defined in the presence of the graviton pole, which is what makes the smeared
gravitational bounds possible.

Appendix~\ref{app:fixed-a-null} rederives the same null constraints from the
crossing-symmetric, or fixed-\(a\), dispersion relation. This representation
goes back to refs.~\cite{Auberson:1972prg,Mahoux:1974ej} and was revived in
the EFT bootstrap by ref.~\cite{Sinha:2020win}. That it automatically contains
the crossing null constraints was pointed out in ref.~\cite{deRham:2022regge};
its power was emphasized in ref.~\cite{Li:2023csdr}, and it has been applied
to gravitational scattering in
refs.~\cite{Chang:2025cxc,Beadle:2025cdx,Pasiecznik:2025eqc}.

For example, the leading moment rows from the \(k=0\) and \(k=2\) dispersion
relations are
\begin{equation}
  c_{0,0}
  =
  \avg{2},
  \qquad
  -c_{1,0}
  =
  \avg{\frac{2}{\mu^2}} .
  \label{eq:first-moment-relations}
\end{equation}
More generally, the \(a^0\) part of the even-\(k\) moment rows gives the
all-order family
\begin{equation}
  c_{m,0}
  =
  (-1)^m
  \avg{\frac{2}{\mu^{2m}}},
  \qquad
  m=0,1,2,\ldots .
  \label{eq:all-cm0-moment-relations-alt}
\end{equation}
In \(D=10\), writing
\begin{equation}
  L_\ell=\ell(\ell+7),
\end{equation}
the first null constraints are
\begin{equation}
  \avg{\frac{L_\ell-2}{2\mu}}=0,
  \qquad
  \avg{
  \frac{L_\ell^2-13L_\ell-20}{20\mu^2}}
  =0 .
  \label{eq:first-null-relations}
\end{equation}
The first null constraint comes from the absence of a term linear in \(t\) on
the low-energy side of the \(k=0\) row. The second comes from subtracting the
two UV moment representations of \(c_{1,0}\).

The gravity-pole row requires a separate treatment, since the massless pole
prevents a naive forward Taylor expansion. One way to handle this contribution
is the smearing method of ref.~\cite{Caron-Huot:2021rmr}. More recently, a
direct fixed-\(t\) implementation was developed in ref.~\cite{Peng:2026ztp}.
Reference~\cite{Peng:2026ztp} reports numerical agreement with the smeared
projective bounds in \(D\ge6\) at the resolutions studied there.

\subsection{Fixed-\texorpdfstring{\(t\)}{t} sampling and finite-dimensional linear program}
\label{subsec:fixed-t-discretization}

We impose the gravity-pole row directly at a finite set of fixed-\(t\) points.
In the numerical implementation we measure all Mandelstam invariants in units
of the gap and set \(M^2=1\). Thus the massive threshold is at \(\mu=1\). For
\(t\in(-1,0)\), we use Chebyshev midpoint nodes
\begin{equation}
  t_i
  =
  -\frac{1}{2}
  +\frac{1}{2}
  \cos\!\left(\frac{(2i-1)\pi}{2N_t}\right),
  \qquad
  i=1,\ldots,N_t .
  \label{eq:fixed-t-cheb-nodes}
\end{equation}
At these points the \(k=-2\) dispersion relation becomes
\begin{equation}
  \sum_{p=1}^{N_\mu}
  \sum_{\substack{0\le \ell\le J_{\max}\\ \ell\ {\rm even}}}
  \rho_{p,\ell}\,
  \mu_p(2\mu_p+t_i)\,
  P_\ell^{(10)}
  \left(1+\frac{2t_i}{\mu_p}\right)
  =
  -\frac{8\pi G}{t_i},
  \qquad
  i=1,\ldots,N_t .
  \label{eq:discrete-pole-row}
\end{equation}
When the gravitational coupling is factored out as an overall normalization, we
may set \(8\pi G=1\). With this convention, the right-hand side reduces to
\(-1/t_i\).

We discretize the high-energy spectral integral and truncate the partial-wave
expansion. The energy variable is compactified by
\begin{equation}
  z=\frac{1}{\mu},
  \qquad
  z\in(0,1],
  \qquad
  \mu=\frac{1}{z}.
  \label{eq:energy-compactification}
\end{equation}
The \(z\)-integral is approximated on the Chebyshev nodes
\begin{equation}
  z_p
  =
  \sin^2\!\left(\frac{p\,\pi}{2N_\mu}\right),
  \qquad
  \mu_p=\frac{1}{z_p},
  \qquad
  p=1,\ldots,N_\mu .
  \label{eq:cheb-energy-nodes}
\end{equation}
The last node is the gap, \(\mu_{N_\mu}=M^2\), so a state exactly at the
cutoff, such as the first string resonance at \(\alpha=1\), lies on the
grid. Doubling \(N_\mu\) keeps every node, so the grids are nested. The
highest energy is \(\mu_1=\csc^2[\pi/(2N_\mu)]\).
The spin sum is truncated to even spins,
\begin{equation}
  \ell=0,2,\ldots,J_{\max}.
  \label{eq:spin-truncation}
\end{equation}

After absorbing the positive quadrature weights, the partial-wave
normalization factors, and the dispersive measure into the variables
\(\rho_{p,\ell}\), the continuous positive spectral measure is replaced by
finitely many non-negative variables,
\begin{equation}
  \rho_\ell(\mu)
  \quad\longrightarrow\quad
  \rho_{p,\ell},
  \qquad
  p=1,\ldots,N_\mu,
  \qquad
  \ell=0,2,\ldots,J_{\max},
  \qquad
  \rho_{p,\ell}\ge0 .
  \label{eq:spectral-discretization}
\end{equation}
Equivalently, each UV average is approximated schematically as
\begin{equation}
  \sum_{\ell=0,2,\ldots}^{\infty}
  n_\ell^{(D)}
  \int_{M^2}^{\infty}
  \frac{\dd \mu}{\pi\mu}\,
  \mu^{2-D/2}
  \rho_\ell(\mu)\,
  (\cdots)
  \quad\longrightarrow\quad
  \sum_{p=1}^{N_\mu}
  \sum_{\substack{0\le \ell\le J_{\max}\\ \ell\ {\rm even}}}
  \rho_{p,\ell}\,
  (\cdots)\big|_{\mu=\mu_p}.
  \label{eq:finite-moment-discretization}
\end{equation}
With this convention, all positive weights have been absorbed into
\(\rho_{p,\ell}\).

The contour factor
\(1/\mu+1/(\mu+t)=\mu^{-1}(2\mu+t)/(\mu+t)\) is written with
\(1/\mu\) in the measure. This factor is included in the positive weights
\(\rho_{p,\ell}\), so the kernel \eqref{eq:fixed-t-kernel} retains its form.

Thus every constraint takes the finite linear form
\begin{equation}
  \sum_{p=1}^{N_\mu}
  \sum_{\substack{0\le \ell\le J_{\max}\\ \ell\ {\rm even}}}
  \rho_{p,\ell}\,
  K_r(\mu_p,\ell)
  =
  R_r(c),
  \qquad
  \rho_{p,\ell}\ge0 .
  \label{eq:finite-lp-row}
\end{equation}
Here \(K_r\) denotes either a fixed-\(t\) gravity-pole kernel, an EFT moment
kernel, or a null-constraint kernel, and \(R_r(c)\) is the corresponding
low-energy side. The numerical bootstrap problem is therefore a
finite-dimensional linear program.

We use this finite LP to obtain the regions shown below. The convergence
checks and the role of high-energy, high-spin states in the sampled graviton
pole are discussed in appendix~\ref{app:revision-convergence}.

\section{Bootstrap with the gravity pole}
\label{sec:gravity-pole}

In this section we keep the massless graviton pole explicitly and use the
fixed-\(t\) dispersive constraints to bound the leading regular coefficient
\(c_{0,0}\). Unless otherwise stated, we use
\begin{equation}
  N_\mu=512,
  \qquad
  J_{\max}=10^4,
  \qquad
  N_t=6,
  \qquad
  N_{\rm null}=55 .
  \label{eq:gravity-baseline}
\end{equation}
Here \(N_\mu\) is the number of energy nodes, \(J_{\max}\) is the spin cutoff
with even spins \(\ell=0,2,\ldots,J_{\max}\), \(N_t\) is the number of
fixed-\(t\) sampling points used for the gravity-pole row, and \(N_{\rm null}\)
is the number of crossing null constraints.

\subsection{Type II string theory as a reference point}
\label{subsec:type-II-reference}

A useful reference point is provided by the tree-level Type II string
amplitude. For both Type IIA and Type IIB string theory, the four-graviton
amplitude is controlled by the Virasoro-Shapiro factor
\begin{equation}
  \M_{\rm VS}(s,t,u)
  =
  -\frac{
  \Gamma(-\alpha' s)\Gamma(-\alpha' t)\Gamma(-\alpha' u)}
  {
  \Gamma(1+\alpha' s)\Gamma(1+\alpha' t)\Gamma(1+\alpha' u)}
  ,
  \qquad
  s+t+u=0 .
  \label{eq:VS-amplitude-alpha-prime}
\end{equation}
This amplitude is meromorphic, crossing symmetric, and has only physical
massive poles. Its low-energy expansion takes the form
\begin{equation}
  \M_{\rm VS}(s,t,u)
  =
  \frac{8\pi G}{stu}
  +
  c_{0,0}
  +
  c_{1,0}x
  +\cdots ,
  \qquad
  x=st+su+tu .
\end{equation}
With the normalization used in this note, the Type II string values are
\begin{equation}
  8\pi G=\frac{1}{\alpha'^3},
  \qquad
  c_{0,0}=2\zeta(3),
  \qquad
  c_{1,0}=-2\alpha'^2\zeta(5).
  \label{eq:type-II-c00-c10}
\end{equation}
The first massive string pole is at \(m^2=1/\alpha'\). Therefore the EFT
expansion is valid below a cutoff \(M\) provided
\begin{equation}
  M^2\le \frac{1}{\alpha'} .
\end{equation}
It is convenient to introduce the dimensionless parameter
\begin{equation}
  \alpha\equiv \alpha' M^2 ,
  \qquad
  0<\alpha\le 1 .
  \label{eq:alpha-dimensionless}
\end{equation}
In terms of this parameter, the Type II string trajectory in the
\(\big(c_{0,0},c_{1,0}\big)\) plane is
\begin{equation}
  \left(
  \frac{c_{0,0}M^6}{8\pi G},
  \frac{c_{1,0}M^{10}}{8\pi G}
  \right)
  =
  \left(
  2\zeta(3)\alpha^3,
  -2\zeta(5)\alpha^5
  \right).
  \label{eq:vs-alpha-line-c00-c10}
\end{equation}
The endpoint \(\alpha=1\) corresponds to placing the first massive string pole
at the cutoff scale. More explicitly, if \(m_1\) denotes the first massive
string mass, then \(\alpha=M^2/m_1^2\) and
\begin{equation}
 \gamma\equiv\frac{c_{0,0}M^6}{8\pi G}
 =2\zeta(3)\left(\frac{M}{m_1}\right)^6 .
 \label{eq:string-normalization-mass-gap}
\end{equation}
Thus the string value \(\gamma=2\zeta(3)\) selects \(m_1=M\) within this
family. This gives a physical interpretation of the normalization used for
our benchmark scans.

\subsection{Allowed region in the \texorpdfstring{\(\big(c_{0,0}M^6/(8\pi G),
c_{1,0}M^{10}/(8\pi G)\big)\)}{(c00 M6/(8 pi G), c10 M10/(8 pi G))} plane}
\label{subsec:c00-c10-region}

We benchmark this linear problem against the ten-dimensional analysis of
ref.~\cite{Albert:2024yap}, which uses smeared functionals and finds
\(c_{0,0}M^6/(8\pi G)\le2.969\) in the same normalization. Here the graviton
pole is instead sampled at fixed \(t\) \cite{Peng:2026ztp} on a finite
positive spectral grid. At the baseline \eqref{eq:gravity-baseline} we find
\(c_{0,0}M^6/(8\pi G)\le2.969\) to four significant digits. This agreement
holds at the stated truncation. We use it as a reference for the nonlinear
scans and do not attempt to determine the optimal gravitational bound.
The numerical checks are given in appendix~\ref{app:revision-convergence}.

We then study the two-dimensional projection onto the first two regular Wilson
coefficients. In this scan we impose the gravity-pole rows, the moment rows, and
the null constraints at the baseline \eqref{eq:gravity-baseline}. We work with the dimensionless
variables
\begin{equation}
  \left(
  \frac{c_{0,0}M^6}{8\pi G},
  \frac{c_{1,0}M^{10}}{8\pi G}
  \right).
  \label{eq:c00-c10-dimensionless-plane}
\end{equation}
The allowed region is reconstructed by optimizing the support function in 192
directions,
\begin{equation}
  \cos\theta\,
  \frac{c_{0,0}M^6}{8\pi G}
  +
  \sin\theta\,
  \frac{c_{1,0}M^{10}}{8\pi G},
\end{equation}
with \(\theta_j=2\pi j/192\), and taking the convex hull of the optima.

\begin{figure}[htbp]
  \centering
  \includegraphics[width=0.68\textwidth]{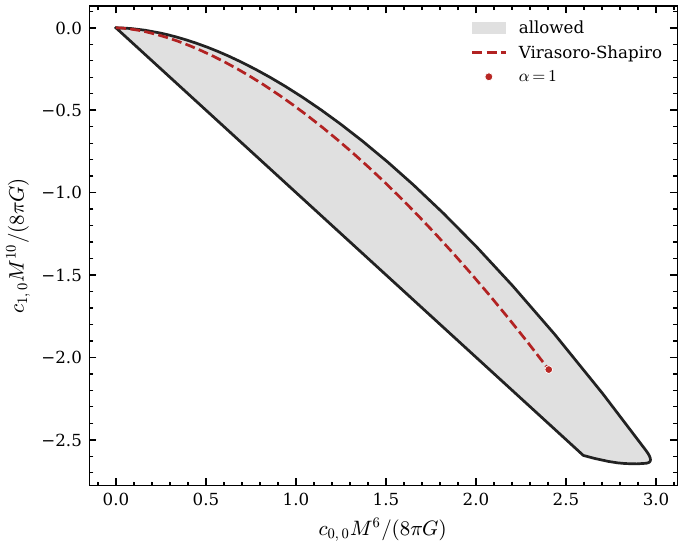}
  \caption{Allowed region in the
  \(\big(c_{0,0}M^6/(8\pi G),c_{1,0}M^{10}/(8\pi G)\big)\) plane at the
  baseline \eqref{eq:gravity-baseline}. The grey region is reconstructed from
  support-function optimization in 192 directions. At this truncation it
  agrees numerically with the region of ref.~\cite{Albert:2024yap}, using
  direct fixed-\(t\) sampling of the graviton pole. The dashed curve is the Type II Virasoro-Shapiro
  trajectory \eqref{eq:vs-alpha-line-c00-c10}, and the red point denotes the
  endpoint \(\alpha=1\).}
  \label{fig:gravity-c00-c10-region}
\end{figure}

\Needspace{4\baselineskip}

\subsection{Extremal spectrum at the maximal value of
\texorpdfstring{\(c_{0,0}M^6/(8\pi G)\)}{c00 M6/(8 pi G)}}
\label{subsec:extremal-spectrum-c00}

\begin{figure}[htbp]
  \centering
  \begin{minipage}{0.48\textwidth}
    \centering
    \includegraphics[width=\textwidth]{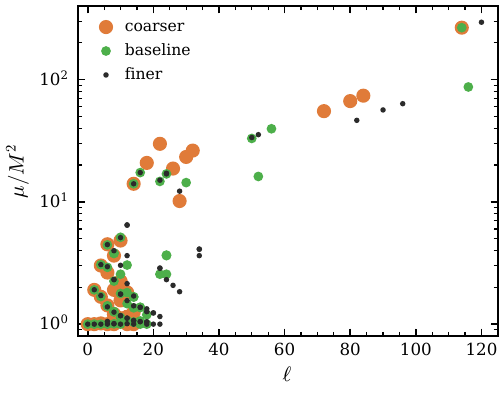}
    \vspace{-2mm}
    \centerline{\small (a) Overview of the extremal support}
  \end{minipage}
  \hfill
  \begin{minipage}{0.48\textwidth}
    \centering
    \includegraphics[width=\textwidth]{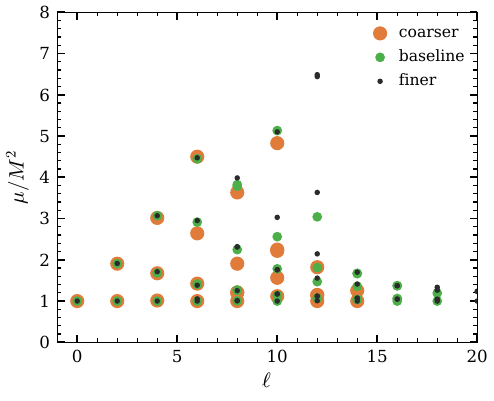}
    \vspace{-2mm}
    \centerline{\small (b) Zoom-in: low-spin and low-mass region}
  \end{minipage}
  \caption{Support of the extremal LP solution at the maximal value of
  \(c_{0,0}M^6/(8\pi G)\), for
  \((N_\mu,J_{\max},N_t,N_{\rm null})=(256,6000,6,36)\) (coarser),
  \((512,10^4,6,55)\) (baseline), and \((1024,16000,6,78)\) (finer).
  Panel (a) gives an overview, and panel (b) magnifies the low-spin,
  low-mass region. A few weights at the highest grid energies lie
  outside the displayed windows.}
  \label{fig:extremal-spectrum-c00}
\end{figure}

We also inspect extremal solutions at the maximal allowed value of
\(c_{0,0}M^6/(8\pi G)\). Since the discretized bootstrap problem is a finite
linear program, the optimizer returns a sparse set of nonzero spectral weights
\(\rho_{p,\ell}\). Figure~\ref{fig:extremal-spectrum-c00} shows weights above the plotting floor
\(\rho_{p,\ell}>10^{-22}\max\rho\) in the \((\ell,\mu)\) plane for three choices of
energy resolution, spin cutoff, and null tower, keeping six gravity samples.

As in the extremal-spectrum analysis of ref.~\cite{Albert:2024yap}, the
discrete support should be interpreted as an extremal solution of the truncated
bootstrap problem, rather than as a unique reconstruction of the physical
spectrum.

\section{Virasoro-inspired ansatz}
\label{sec:virasoro-inspired-ansatz}

\subsection{Motivation}
\label{subsec:virasoro-inspired-motivation}

The closed-string Virasoro-Shapiro amplitude admits an exponentiated
low-energy expansion. After factoring out the overall gravitational coupling
and working in string units, \(8\pi G=1/\alpha'^3=1\), it can be written as
\cite{Geiser:2022exp}

\begin{equation}
  \M_{\rm VS}(s,t,u)
  =
  \frac{1}{stu}
  \exp\left[
  \sum_{k\ge1}
  \frac{2\zeta(2k+1)}{2k+1}
  \left(
  s^{2k+1}+t^{2k+1}+u^{2k+1}
  \right)
  \right],
  \qquad
  s+t+u=0 .
  \label{eq:vs-exponential-form}
\end{equation}
Equivalently, after removing the universal gravity pole by defining
\begin{equation}
  B(s,t,u)
  \equiv
  \frac{stu\,\M(s,t,u)}{8\pi G},
  \label{eq:def-B}
\end{equation}
the exact Virasoro-Shapiro amplitude satisfies
\begin{equation}
  \log B_{\rm VS}(s,t,u)
  =
  \sum_{k\ge1}
  \frac{2\zeta(2k+1)}{2k+1}
  \left(
  s^{2k+1}+t^{2k+1}+u^{2k+1}
  \right).
  \label{eq:log-b-vs}
\end{equation}
Thus the primitive data in the logarithm are fully crossing-symmetric
odd-power combinations.

A similar structure appears in the four-dimensional planar open-string
setup of ref.~\cite{Elvang:2026maxsusy}: maximal supersymmetry, $SU(4)$
R-symmetry, higher-point tree-level factorization, and a parity condition on
the six-scalar amplitudes imply constraints on the four-point EFT coefficients. At the EFT orders tested, the resulting constraints are compatible
with the resummed parametrization
\begin{equation}
  A_4[--++]
  =
  \langle 12\rangle^2[34]^2\,F(s,t),
  \label{eq:A4-open-F}
\end{equation}
where
\begin{equation}
  F(s,t)
  =
  F_0(s,t)
  \exp\left[
  \sum_{k\ge1}
  \frac{\alpha_{2k+1,0}}{2k+1}
  \left(
  s^{2k+1}+t^{2k+1}+u^{2k+1}
  \right)
  \right].
  \label{eq:open-string-exponential-motivation}
\end{equation}
Setting the scale parameter in \(F_0\) to one, the symmetric prefactor can be
written as
\begin{equation}
  F_0(s,t)
  =
  -\frac{1}{st}
  \sqrt{
  \frac{
  \pi st\,\sin\!\left[\pi(s+t)\right]}
  {
  (s+t)\,\sin(\pi s)\,\sin(\pi t)}
  } .
  \label{eq:F0-open-string}
\end{equation}
Extending this pattern to all orders is additional input. The computed
relations provide motivation for exponentiation, rather than a derivation of
the complete ten-dimensional ansatz.

The open-string structure is naturally related to the closed-string
Virasoro-Shapiro form through the Kawai--Lewellen--Tye relation
\cite{Kawai:1985xq}. At four points, KLT expresses the closed-string amplitude
as a bilinear in open-string amplitudes. For the string scalar factors, after
removing the polarization factors and expressing both open factors in the same
ordering, the normalization $g=8\pi G$ gives
\begin{equation}
  \M_{\rm closed}(s,t,u)
  =
  -g F(s,t)\,
  S_{\rm KLT}(s,t)\,
  F(s,t),
  \label{eq:klt-four-point}
\end{equation}
with the explicit sine kernel
\begin{equation}
  S_{\rm KLT}(s,t)
  =
  \frac{\sin(\pi s)\sin(\pi t)}
       {\pi\sin\!\left[\pi(s+t)\right]} .
  \label{eq:klt-kernel}
\end{equation}
The overall sign matches the gravity-pole convention of this note:
$-g F_0(s,t)^2S_{\rm KLT}(s,t)=g/(stu)$ at the chosen unit scale.
Thus the string amplitude is obtained from the squared open scalar factor
together with the sine kernel and the stated coupling normalization.
This motivates a closed-string ansatz in which the logarithm of the
pole-stripped amplitude is built from crossing-symmetric primitive structures.

\subsection{The ansatz and induced low-energy constraints}
\label{subsec:virasoro-inspired-ansatz}

Motivated by the Virasoro-Shapiro form and the higher-point/KLT structure
described above, we impose a string-inspired ansatz on the low-energy EFT data.
More precisely, we do not assume that the full amplitude is exactly of
Virasoro-Shapiro form. We only assume that the pole-subtracted low-energy
expansion can be organized as
\begin{equation}
  \log B(s,t,u)
  =
  \sum_{k\ge1}
  \alpha_{2k+1}
  \left(
  s^{2k+1}+t^{2k+1}+u^{2k+1}
  \right),
  \qquad
  s+t+u=0 .
  \label{eq:virasoro-inspired-ansatz}
\end{equation}
We call this the Virasoro-inspired ansatz. The coefficients
\(\alpha_{2k+1}\) are kept free in the bootstrap. Thus the ansatz constrains
only the pattern of low-energy Wilson coefficients, rather than fixing them to
their string-theory values. For the exact Virasoro-Shapiro amplitude,
\begin{equation}
  \alpha_{2k+1}
  =
  \frac{2\zeta(2k+1)}{2k+1}.
  \label{eq:vs-alpha-values}
\end{equation}

The ansatz \eqref{eq:virasoro-inspired-ansatz} is additional input; locality
alone does not imply it. Local crossing-symmetric data only require
\(\log B=yH(x,y)\) with \(H\) analytic at the origin, which allows, for
example, \(y^2\) at degree six, or products of power sums such as
\(p_3^2=9y^2\). Thus
\(\log B=\alpha_3(s^3+t^3+u^3)+\beta(stu)^2\) is local, but it gives
\(c_{1,1}=c_{0,0}^2/2+\beta\) in \(8\pi G=1\) units and violates NL1 below
when \(\beta\neq0\). The assumption that removes such structures is channel
factorization of the pole-stripped amplitude,
\begin{equation}
 B(s,t,u)=h(s)\,h(t)\,h(u),
 \label{eq:channel-factorization}
\end{equation}
with \(\log h\) analytic at the origin. Locality then removes the even part of
\(\log h\), as shown below and in appendix~\ref{app:power-sums-locality}, so
that \(\log h(z)=\sum_{k\ge1}\alpha_{2k+1}z^{2k+1}\) and
\eqref{eq:channel-factorization} reproduces
\eqref{eq:virasoro-inspired-ansatz}. For the Virasoro-Shapiro amplitude
\begin{equation}
 h_{\rm VS}(z)=e^{-2\gamma_E z}\,\frac{\Gamma(1-z)}{\Gamma(1+z)},
 \label{eq:vs-channel-factor}
\end{equation}
where the exponential factors cancel in the product because \(s+t+u=0\). We
regard \eqref{eq:channel-factorization} as a tree-level hypothesis, which the
Virasoro-Shapiro amplitude obeys; beyond tree level, massless loops produce
non-analytic terms that violate it. We do not claim that generic UV
completions obey it. Since every \(\alpha_{2k+1}\) is fixed by a forward
coefficient, see \eqref{eq:linear-alpha-all-order} below, the ansatz
reconstructs the full low-energy amplitude from its forward limit. The
relations NL1--NL6 below are the first of these predictions, and the bootstrap
asks whether the predicted non-forward data admit a positive partial-wave
expansion. This is why they are much stronger than analyticity, crossing, and
unitarity.

Within the assumed linear combination of single-variable power sums, locality
requires the absence of even powers. In the following algebra we use
\(8\pi G=1\). Indeed,
\begin{equation}
  \M(s,t,u)
  =
  \frac{B(s,t,u)}{stu}
  =
  \frac{B(x,a)}{xa},
  \qquad
  x=st+su+tu,
  \qquad
  xa=stu .
\end{equation}
The gravity pole is the single term \(1/(stu)\). Therefore \(B-1\) must be
divisible by \(stu=xa\), so that the remaining part of \(\M\) is a local
polynomial. Odd power sums obey this condition, for instance
\begin{equation}
  s^3+t^3+u^3=3stu .
\end{equation}
Even power sums do not. For example,
\begin{equation}
  s^2+t^2+u^2=-2x
\end{equation}
would generate
\begin{equation}
  B=1-2\beta_2x+\cdots,
  \qquad
  \M=\frac{1}{xa}-\frac{2\beta_2}{a}+\cdots ,
\end{equation}
where the second term is an additional nonlocal singularity beyond the gravity
pole. We give an all-order proof in appendix~\ref{app:power-sums-locality}.

The linear terms in the \(\alpha\)'s can be written to all orders. Define
\[
  p_n=s^n+t^n+u^n,
  \qquad
  x=st+su+tu,
  \qquad
  y=stu=xa,
  \qquad
  s+t+u=0 .
\]
Newton's identities imply
\begin{equation}
  p_n=-x\,p_{n-2}+y\,p_{n-3}.
  \label{eq:power-sum-recursion-linear-alpha}
\end{equation}
For odd powers this gives
\begin{equation}
  p_{2k+1}
  =
  (2k+1)(-1)^{k+1}x^{k-1}y
  +O(y^2),
  \qquad k\ge1 .
  \label{eq:odd-power-leading-y}
\end{equation}
Since \(y=xa\), we have
\begin{equation}
  p_{2k+1}
  =
  (2k+1)(-1)^{k+1}x^k a
  +O(a^2).
\end{equation}
The ansatz
\begin{equation}
  \log B
  =
  \sum_{k\ge1}\alpha_{2k+1}p_{2k+1}
\end{equation}
therefore contains, at linear order in the \(\alpha\)'s,
\begin{equation}
  \log B
  =
  \sum_{k\ge1}
  (2k+1)(-1)^{k+1}\alpha_{2k+1}x^k a
  +O(a^2).
\end{equation}
Because \(\M=B/(xa)\), these terms contribute to the \(a^0\) Wilson
coefficients as
\begin{equation}
  c_{k-1,0}
  =
  (2k+1)(-1)^{k+1}\alpha_{2k+1},
  \qquad k\ge1 .
  \label{eq:linear-alpha-all-order}
\end{equation}
Equivalently, writing \(m=k-1\),
\begin{equation}
 c_{m,0}
  =
  (2m+3)(-1)^m\alpha_{2m+3},
  \qquad m\ge0 .
  \label{eq:linear-alpha-all-order-m}
\end{equation}
The first few cases are
\begin{equation}
  c_{0,0}=3\alpha_3,
  \qquad
  c_{1,0}=-5\alpha_5,
  \qquad
  c_{2,0}=7\alpha_7,
  \qquad
  c_{3,0}=-9\alpha_9 .
\end{equation}
The exponential form also generates nonlinear relations among the Wilson
coefficients. We impose these relations level by level. The first six
nonlinear levels are
\begin{align}
  {\rm NL1}:&\qquad
  c_{1,1}
  =
  \frac12 c_{0,0}^2,
  \label{eq:ansatz-NL1}
  \\
  {\rm NL2}:&\qquad
  c_{2,1}
  =
  c_{0,0}c_{1,0},
  \label{eq:ansatz-NL2}
  \\
  {\rm NL3}:&\qquad
  c_{2,2}
  +\frac13 c_{3,0}
  =
  \frac16 c_{0,0}^3,
  \label{eq:ansatz-NL3}
  \\
  {\rm NL4}:&\qquad
  c_{3,1}
  =
  c_{0,0}c_{2,0}
  +\frac12 c_{1,0}^2,
  \label{eq:ansatz-NL4}
  \\
  {\rm NL5}:&\qquad
  c_{3,2}
  =
  -c_{4,0}
  +\frac12 c_{0,0}^2c_{1,0},
  \label{eq:ansatz-NL5}
  \\
  {\rm NL6}:&\qquad
  c_{3,3}
  =
  -\frac13 c_{0,0}c_{3,0}
  +\frac{1}{24}c_{0,0}^4 .
  \label{eq:ansatz-NL6}
\end{align}
Here ``NL\(n\)'' denotes the \(n\)-th nonlinear truncation level of the
Virasoro-inspired ansatz. For example, an NL3 scan imposes
\eqref{eq:ansatz-NL1}--\eqref{eq:ansatz-NL3}, while an NL6 scan imposes
\eqref{eq:ansatz-NL1}--\eqref{eq:ansatz-NL6}.

The same expansion can be continued systematically. These higher nonlinear
relations are generated by expanding \(\exp(\log B)\), using
\(\M=B/(xa)\), and eliminating the free parameters \(\alpha_{2k+1}\) in favor
of the linear coefficients \(c_{m,0}\). These relations are not consequences
of four-point positivity alone. They are additional string-inspired constraints
motivated by the exponentiated Virasoro-Shapiro structure. In the numerical
bootstrap we impose them level by level and test whether the allowed region is
driven toward the
Virasoro-Shapiro point.

In the finite NL3 and NL6 scans the ansatz is used only to generate
polynomial relations among finitely many Wilson coefficients. Each LP
imposes a finite set of sum rules derived under the
Regge assumption; a feasible measure need not reconstruct an amplitude
with that high-energy behaviour. If the ansatz is
instead imposed to all orders, the forward sum rules build \(\log B\) from a
positive measure (section~\ref{subsec:stripped-analytic-structure}), and its
growth at fixed \(t<0\) is controlled by the large-mass density of that
measure (appendix~\ref{app:positivity-regge}). Regge boundedness therefore
constrains only the tail of the measure. It does not constrain the low-lying
levels that dominate the \(\alpha_{2k+1}\), and by itself it does not single
out the Virasoro-Shapiro values.

\section{Gravity-pole bootstrap with nonlinear conditions}
\label{subsec:string-line}
\subsection{Numerical implementation of the nonlinear ansatz}
\label{subsec:numerical-nonlinear-ansatz}

We now impose the nonlinear Virasoro-inspired relations introduced above. At
this stage the bootstrap problem is no longer a linear program if all Wilson
coefficients are varied simultaneously. A direct nonlinear optimization would
therefore lead to a non-convex problem. Instead, we use a feasibility-scan
strategy: we fix some of the lower Wilson coefficients and then check whether
the remaining constraints admit a positive spectral density.

In this section we use the baseline \eqref{eq:gravity-baseline}. At NL3,
fixing \(c_{0,0}\) makes \eqref{eq:ansatz-NL1}--\eqref{eq:ansatz-NL3}
linear, so the ends of each
\(c_{1,0}\) slice follow from two LP optimizations. At NL6 we also fix a trial
\(c_{1,0}\), which also fixes the squared term in \eqref{eq:ansatz-NL4};
the remaining problem is an LP feasibility test. The main cost
is then locating the boundary of the narrow feasible windows. We first
scan the NL3 interval, retain verified feasible and excluded points, and
bisect every identified transition until its bracket reaches the requested
resolution. Resolving a bracket of width \(\Delta\) to \(\epsilon\) requires
approximately \(\lceil\log_2(\Delta/\epsilon)\rceil\) successive LP tests.
Thus the higher-NL search can require many solves even though each
fixed-coefficient subproblem is linear. The precise scan, bisection, and
acceptance rules are given in appendix~\ref{app:numerical-details}.

We compare the bounds with the Virasoro--Shapiro trajectory
\eqref{eq:vs-alpha-line-c00-c10}.
\subsection{Results for the allowed region}
\label{subsec:nonlinear-results}

\begin{figure}[htbp]
  \centering
  \begin{minipage}{0.48\textwidth}
    \centering
    \includegraphics[width=\textwidth]{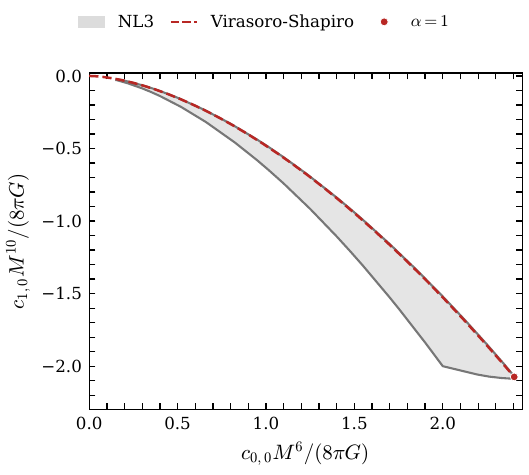}
    \vspace{-2mm}
    \centerline{\small (a) NL3 allowed region}
  \end{minipage}
  \hfill
  \begin{minipage}{0.48\textwidth}
    \centering
    \includegraphics[width=\textwidth]{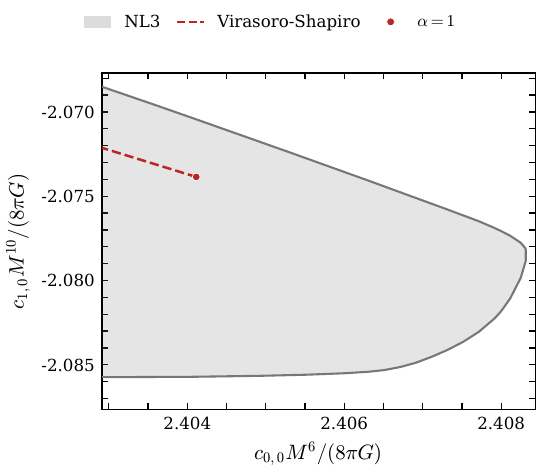}
    \vspace{-2mm}
    \centerline{\small (b) Zoom near \(\alpha=1\)}
  \end{minipage}
  \caption{Allowed region after imposing NL1--NL3 at the baseline
  \eqref{eq:gravity-baseline}. The shaded region joins the minima and maxima of
  $c_{1,0}$ at fixed $c_{0,0}$. The dashed curve is the Virasoro--Shapiro trajectory, and the
  red point denotes the endpoint $\alpha=1$. Panel (b) magnifies the
  neighbourhood of $\alpha=1$. Both panels show the coefficient values directly.}
  \label{fig:NL3-region}
\end{figure}

We first impose the first three nonlinear relations, NL1--NL3, and scan the
allowed region in the
\(\big(c_{0,0}M^6/(8\pi G),c_{1,0}M^{10}/(8\pi G)\big)\) plane. The result is
shown in figure~\ref{fig:NL3-region}. The nonlinear constraints already shrink
the allowed region significantly. The upper boundary of the allowed band lies
close to the Virasoro--Shapiro trajectory. At the string reference slice
\(\gamma\equiv c_{0,0}M^6/(8\pi G)=2\zeta(3)\) the NL3 interval is
\(c_{1,0}M^{10}/(8\pi G)+2\zeta(5)\in[-1.185\times10^{-2},3.415\times10^{-3}]\),
which contains the string value at zero offset. The region extends slightly
beyond the string endpoint \(\alpha=1\).

\begin{figure}[htbp]
  \centering
  \begin{minipage}{0.48\textwidth}
    \centering
    \includegraphics[width=\textwidth]{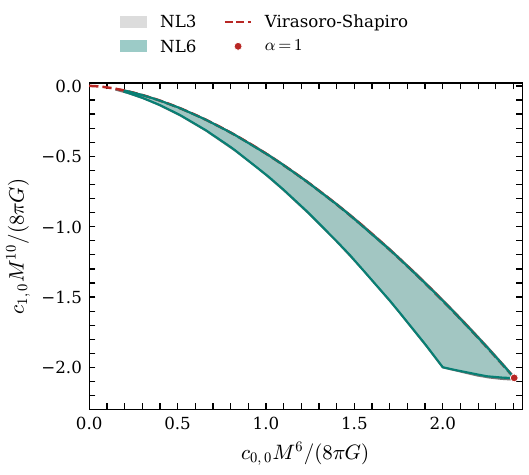}
    \vspace{-2mm}
    \centerline{\small (a) Comparison of NL3 and NL6}
  \end{minipage}
  \hfill
  \begin{minipage}{0.48\textwidth}
    \centering
    \includegraphics[width=\textwidth]{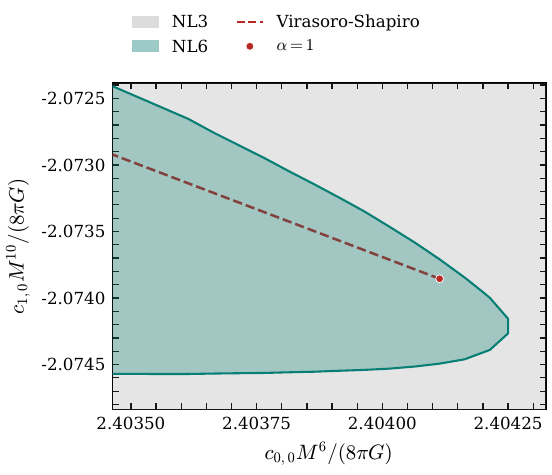}
    \vspace{-2mm}
    \centerline{\small (b) Zoom near $\alpha=1$}
  \end{minipage}
  \caption{Comparison of the NL3 and NL6 regions at the settings of
  figure~\ref{fig:NL3-region}. Light shading: NL3 region; teal shading: range
  of the feasible NL6 tests at fixed $c_{0,0}$, with adjacent slices joined
  for display. Short end segments close the outline at the extreme checked
  slices; they do not locate the continuous NL6 boundary. The
  dashed curve and the red point are as in figure~\ref{fig:NL3-region}. Panel (b)
  magnifies the neighbourhood of $\alpha=1$. Both panels show the
  coefficient values directly.}
  \label{fig:NL3-NL6-region}
\end{figure}

We then impose the next three nonlinear relations, NL4--NL6. The comparison
between the NL3 and NL6 scans is shown in figure~\ref{fig:NL3-NL6-region}.
At \(\gamma=2\zeta(3)\) the feasible NL6 tests span
\(c_{1,0}M^{10}/(8\pi G)+2\zeta(5)\in[-6.377,1.461]\times10^{-4}\),
a window of width \(7.838\times10^{-4}\) that contains the string value. The window closes
rapidly near the string endpoint.

These results show that the Virasoro-inspired relations strongly constrain
the allowed EFT data and drive the allowed region toward the Virasoro-Shapiro
trajectory. They are results at the stated truncation. As discussed in
appendix~\ref{app:revision-convergence}, sampling the pole at finitely many
\(t\) does not control the high-energy, high-spin region in which the pole
is built. On the string slice, however, the NL6 window depends little on the
sampled pole: removing the pole rows changes its width by only \(0.6\%\).
The pole-subtracted problem of section~\ref{sec:stripped} has no sampled pole
row, and doubling its energy or spin resolution changes the NL6 island by at
most \(0.1\%\).

\section{Gravity-pole-subtracted bootstrap with nonlinear conditions}
\label{sec:stripped}
We now take a step back and consider a gravity-pole-subtracted setup. In this
case the massless graviton pole is removed before imposing the bootstrap
constraints. Although this is less direct than the full gravitational bootstrap
considered above, it provides a useful complementary perspective and makes the
regular part of the amplitude easier to analyze.

We define the pole-subtracted amplitude by
\begin{equation}
  \M_{\rm sub}(s,t,u)
  \equiv
  \M(s,t,u)-\frac{8\pi G}{stu}
  =
  8\pi G\,\frac{B(s,t,u)-1}{stu}.
  \label{eq:pole-subtracted-amplitude}
\end{equation}
The nonlinear relations introduced above involve the gravitational coupling:
restoring \(g=8\pi G\), for example, NL1 reads
\(c_{1,1}=c_{0,0}^{\,2}/(2g)\). They therefore fix the strength of the
local interactions relative to graviton exchange. Below we use Wilson
coefficients normalized by \(g\) and suppress this common factor.

\subsection{Numerical result}
\label{subsec:stripped-numerics}

We first perform a numerical bootstrap scan in the pole-subtracted setup.
Without the nonlinear relations this problem is homogeneous in the spectral
measure, and \(c_{0,0}\) would only set the overall normalization of the
regular part of the amplitude. The nonlinear relations are not homogeneous,
so \(c_{0,0}\) becomes a physical slice parameter. We fix it to the
Virasoro-Shapiro value,
\begin{equation}
  c_{0,0}=2\zeta(3),
\end{equation}
and study the allowed region of the two dimensionless ratios
\begin{equation}
  \left(
  \frac{M^4 c_{1,0}}{c_{0,0}},
  \frac{M^8 c_{2,0}}{c_{0,0}}
  \right).
\end{equation}
We choose this slice to test how much spectral information follows from the
nonlinear relations. On the string family it places the first massive state
at \(M\), as explained in \eqref{eq:string-normalization-mass-gap}.
By NL1, \eqref{eq:ansatz-NL1}, it also fixes
\(M^6c_{1,1}/c_{0,0}=\gamma/2=\zeta(3)\). This is the string value of
the ratio \(g_3/g_0\), in mass-gap units, scanned in the four-dimensional
bootstrap of ref.~\cite[section~6.2 and figure~1]{Berman:2026gravityparity}.
There the nonlinear relations generate a corner approaching the string
point, providing further motivation for this slice.
Our numerical states and residues at \(\mu=1\) remain free; related
open-string studies instead use explicit low-mass spectral input
\cite{Berman:2024wyt,Berman:2025owb}. The analytic derivation in
section~\ref{subsec:stripped-analytic-structure} uses its stated all-order
and spectral hypotheses.

How the region depends on this choice follows from a scaling property. Because
the nonlinear relations contain \(g\), they fix the normalization of the
spectral measure relative to graviton exchange, and \(\gamma\) measures the
position of the spectrum relative to the cutoff. The rescaling
\(B(s,t,u)\to B(\lambda s,\lambda t,\lambda u)\) at fixed \(g\) preserves
the ansatz \eqref{eq:virasoro-inspired-ansatz}, with
\(\alpha_{2k+1}\to\lambda^{2k+1}\alpha_{2k+1}\), and maps
\(c_{m,n}\to\lambda^{3+2m+n}c_{m,n}\). Every nonlinear relation is therefore
invariant. On the dispersive side the same map replaces
\(\rho_\ell(\mu)\) by \(\lambda^{5-D/2}\rho_\ell(\lambda\mu)\), which moves
each state from \(\mu\) to \(\mu/\lambda\) and preserves positivity, the sum
rules and the null constraints. For \(\lambda\le1\) the moved spectrum lies
above the gap, so every allowed point at \(\gamma\) gives an allowed point at
\(\lambda^3\gamma\). In the scale-invariant ratios
\(c_{1,0}/c_{0,0}^{5/3}\) and \(c_{2,0}/c_{0,0}^{7/3}\), the allowed region
can therefore only grow as \(\gamma\) decreases. Lowering \(\gamma\) is
equivalent to raising the spectrum above the cutoff, which relaxes the gap
condition. Among the slices that contain the Virasoro-Shapiro point,
\(\gamma\le2\zeta(3)\), the string slice is the most restrictive, and the
narrow island is specific to it.

The numerical setup is otherwise as above, except that we do not impose the
\(k=-2\) gravity-pole row. In this section we use
\begin{equation}
  N_\mu=512,
  \qquad
  J_{\max}=2000,
  \qquad
  N_{\rm null}=78 .
  \label{eq:pole-baseline}
\end{equation}

\begin{figure}[t]
  \centering
  \includegraphics[width=0.90\textwidth]{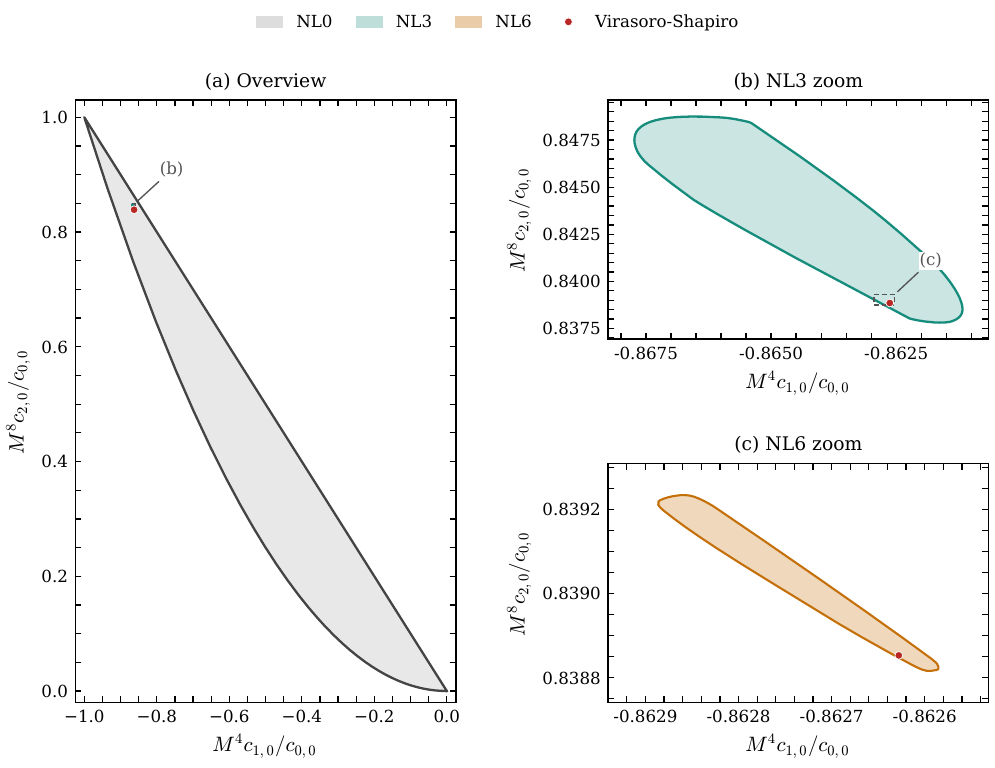}
  \caption{Pole-subtracted allowed regions at \(c_{0,0}=2\zeta(3)\) and the
  baseline \eqref{eq:pole-baseline}. Grey and teal: NL0 and NL3 regions,
  reconstructed from support-function optima. Orange: NL6 island,
  joining the \(c_{2,0}\) intervals at fixed \(c_{1,0}\), with the outline
  closed at the extreme checked slices. The red point is the
  Virasoro--Shapiro point \((-\zeta(5)/\zeta(3),\zeta(7)/\zeta(3))\). Panels
  (b) and (c) show the whole NL3 contour and displayed NL6 island. Dashed
  boxes in (a) and (b) mark the extents of (b) and (c), respectively.
  All panels show the coefficient ratios directly.}
  \label{fig:pole-subtracted-nl036}
\end{figure}

We then impose the Virasoro-inspired nonlinear relations at different
truncation levels. We denote by NL0 the purely linear bootstrap with no
nonlinear relations imposed, by NL3 the scan with the first three nonlinear
relations, and by NL6 the scan with the first six nonlinear relations. The
result is shown in fig.~\ref{fig:pole-subtracted-nl036}. As the nonlinear
relations are added, the allowed region shrinks dramatically from a broad
linear region to a small island containing the Virasoro-Shapiro point. At the
string slice \(c_{1,0}=-2\zeta(5)\), the NL6 interval is
\(c_{2,0}-2\zeta(7)\in[-13.45,+116.3]\times10^{-6}\).

The regions are shown at the settings \eqref{eq:pole-baseline}; their
dependence on the numerical truncations is discussed in
appendix~\ref{app:revision-convergence}.

It is also useful to examine the extremal spectrum at a boundary point near
the Virasoro-Shapiro solution. At the NL6 level, after imposing the nonlinear
relations and fixing the lower Wilson coefficients, the bootstrap problem
is linear in the spectral weights, with \(c_{2,0}\) as the objective. We
fix \(c_{1,0}\) to its Virasoro-Shapiro value and minimize
\(c_{2,0}\). The resulting solution corresponds to a boundary point close to
the string point.

\begin{figure}[t]
  \centering
  \begin{minipage}{0.48\textwidth}
    \centering
    \includegraphics[width=\textwidth]{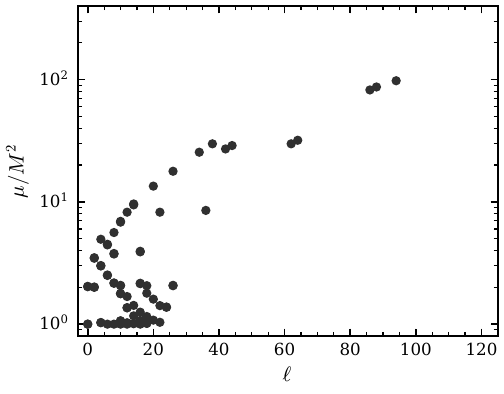}
    \vspace{-2mm}
    \centerline{\small (a) Overview of the extremal support}
  \end{minipage}
  \hfill
  \begin{minipage}{0.48\textwidth}
    \centering
    \includegraphics[width=\textwidth]{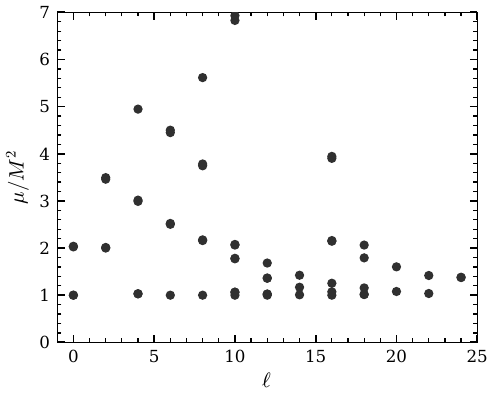}
    \vspace{-2mm}
    \centerline{\small (b) Zoom-in: low-spin and low-mass region}
  \end{minipage}
  \caption{Extremal spectrum for the NL6 pole-subtracted bootstrap at the
  boundary point obtained by fixing \(c_{1,0}\) to the Virasoro-Shapiro value
  and minimizing \(c_{2,0}\), at the baseline \eqref{eq:pole-baseline}.
  Panel (a) gives an overview, and panel (b) magnifies the low-spin,
  low-mass region. The few weights at \(\mu\gtrsim7\times10^3M^2\)
  lie outside the displayed windows.}
  \label{fig:pole-subtracted-nl6-spectrum}
\end{figure}

\paragraph{Spectrum of the boundary point near the string point at NL6.}
The support of the corresponding extremal spectrum is shown in
fig.~\ref{fig:pole-subtracted-nl6-spectrum}. As in the previous
extremal-spectrum analysis, this discrete support should be interpreted as an
extremal solution of the truncated bootstrap problem, rather than as a unique
reconstruction of the physical string spectrum. The left panel shows the
support in the \((\ell,\mu)\) plane, while the right panel zooms into the
low-spin and low-mass region.

The upper edge of the support rises roughly linearly with \(\ell\).
It is nevertheless not the string spectrum;
for example, it has weight at the gap \(\mu=M^2\) for spins up to
\(\ell=22\). Sparsity is a generic feature of a finite LP optimizer, and the
plot is not evidence that the physical measure is discrete, that the logarithmic
weights are one, or that finitely many spins occur at a mass level. Their roles
are distinguished in the conditional analytic argument below.

\Needspace{18\baselineskip}
\subsection{Analytic origin of the string spectrum}
\label{subsec:stripped-analytic-structure}

The all-order ansatz relates the amplitude to its regular forward limit.
We now show how positivity fixes the logarithmic weights of isolated
exchanges, and how their spectrum follows from a finite number of spins
at the first massive pole. The argument uses the following assumptions:
\begin{itemize}
\item[(H1)] the all-order ansatz \eqref{eq:virasoro-inspired-ansatz};
\item[(H2)] the forward dispersion relations with a positive spectral density;
\item[(H3)] partial-wave positivity \eqref{eq:partial-wave} at fixed
  momentum transfer \(-s\le t<0\);
\item[(H4)] isolated massive exchanges with no finite accumulation of masses;
\item[(H5)] finitely many spins at the first massive level.
\end{itemize}
The last two are spectral assumptions. Unit logarithmic weights will follow
from positivity rather than enter as a separate assumption.

\paragraph{From forward moments to a product.}
Write \(g=8\pi G\), \(B=stu\,\M/g\), and
\(\widehat c_{m,n}=c_{m,n}/g\), and let \(M^2\) be the mass gap. The all-order
ansatz is
\begin{equation}
 \log B=\sum_{k\ge1}\alpha_{2k+1}
 (s^{2k+1}+t^{2k+1}+u^{2k+1}),\qquad s+t+u=0.
 \label{eq:odd-logB-moment-start}
\end{equation}
After subtracting the massless pole, the bracket
\eqref{eq:bracket-definition} defines a positive measure on the massive
forward spectrum,
\begin{equation}
 d\omega(\sigma)=\sum_\ell n_\ell^{(D)}
 \frac{d\sigma}{\pi\sigma}\sigma^{2-D/2}\rho_\ell(\sigma)
 =\frac{\operatorname{Im}\M_{\rm sub}(\sigma,0)}{\pi\sigma}\,d\sigma.
 \label{eq:forward-omega-measure}
\end{equation}
This includes delta-function contributions from isolated exchanges. The
forward sum rules and the relation between the odd logarithmic coefficients
and \(\widehat c_{m,0}\) give
\begin{equation}
 \alpha_{2m+3}=\frac{(-1)^m\widehat c_{m,0}}{2m+3}
 =\frac{2}{g(2m+3)}\int_{M^2}^{\infty}
 \sigma^{-2m}\,d\omega(\sigma),\qquad m\ge0.
\end{equation}
It is useful to absorb the mass factors into the positive measure
\begin{equation}
 d\nu(\sigma)\equiv\frac{\sigma^3}{g}\,d\omega(\sigma)
 =\frac{\sigma^2}{\pi g}\operatorname{Im}\M_{\rm sub}(\sigma,0)\,d\sigma,
 \qquad d\nu\ge0,
 \label{eq:logarithmic-nu-measure}
\end{equation}
so that the logarithmic coefficients are inverse moments,
\begin{equation}
 \alpha_{2k+1}=\frac{2}{2k+1}\int_{M^2}^{\infty}
 \frac{d\nu(\sigma)}{\sigma^{2k+1}},\qquad k\ge1.
 \label{eq:alpha-odd-moment-representation}
\end{equation}
Finite \(c_{0,0}\) therefore gives
\(\int\sigma^{-3}d\nu=c_{0,0}/(2g)<\infty\).

These moments allow us to resum the ansatz in the low-energy domain
\(\max(|s|,|t|,|u|)<M^2\). We use the expansion
\begin{equation}
 \log\frac{\sigma+z}{\sigma-z}
 =2\sum_{q\ge0}\frac{z^{2q+1}}{(2q+1)\sigma^{2q+1}},
 \qquad |z|<\sigma,
 \label{eq:log-ratio-odd-expansion}
\end{equation}
with the branch analytic at \(z=0\). In the sum over \(s,t,u\), the linear
terms cancel and the combined integrand is \(O(\sigma^{-3})\). The finite
third inverse moment then justifies exchanging the series and integral on
compact subsets of the low-energy domain. This gives
\begin{equation}
 \log B=\int_{M^2}^{\infty}d\nu(\sigma)
 \log\left[\frac{(\sigma+s)(\sigma+t)(\sigma+u)}
 {(\sigma-s)(\sigma-t)(\sigma-u)}\right].
 \label{eq:closed-logB-moment-form}
\end{equation}
The representation follows from the ansatz and the forward sum rules. Its
branch is fixed by continuation from the origin; the branch points of the
individual logarithms lie at \(z=\pm\sigma\), outside the low-energy domain.
A continuous measure is still allowed at this stage.

For isolated massive exchanges, assumption (H4) makes the measure discrete:
\begin{equation}
 d\nu(\sigma)=\sum_n w_n\delta(\sigma-\sigma_n)\,d\sigma,
 \quad w_n>0,\quad 0<\sigma_1<\sigma_2<\cdots,
 \quad\sum_n\frac{w_n}{\sigma_n^3}<\infty.
 \label{eq:discrete-measure}
\end{equation}
Here \(\sigma_n\) are the squared masses of the exchanged levels.
The grouped logarithms therefore exponentiate to the product
\begin{equation}
 B=\prod_n\left[
 \frac{(\sigma_n+s)(\sigma_n+t)(\sigma_n+u)}
 {(\sigma_n-s)(\sigma_n-t)(\sigma_n-u)}\right]^{w_n}.
 \label{eq:closed-B-discrete-product}
\end{equation}
Each three-channel factor differs from one by \(O(\sigma_n^{-3})\).
The summability condition thus gives local uniform convergence on an
analytic branch away from the singularities. Keeping the three channels
together is essential for this convergence.

\paragraph{Fixing the weights.}
Forward positivity gives \(w_n>0\). To determine their values, we use
partial-wave positivity away from the forward limit. Since
\(n_\ell^{(D)}\rho_\ell\ge0\) and \(|P_\ell^{(D)}(x)|\le1\) for \(|x|\le1\),
the partial-wave expansion \eqref{eq:partial-wave} gives
\begin{equation}
 |\operatorname{Im}\M(s,t)|\le\operatorname{Im}\M(s,0),\qquad -s\le t\le0,
 \label{eq:nonforward-bound}
\end{equation}
as measures in \(s\). Hence a non-forward discontinuity can have no support
where the forward spectral measure vanishes. For \(|t|<\sigma_1\), every
\(\log[(\sigma_n+t)/(\sigma_n-t)]\) is real, and
\(\operatorname{Im}\log[(\sigma+z+i0)/(\sigma-z-i0)]=\pi\theta(z-\sigma)\) for
\(z>0\). Hence, for \(t<0\),
\begin{equation}
 \operatorname{Im}\log B(s+i0,t)=\pi\!\!\sum_{n:\,s+t<\sigma_n<s}\!\! w_n .
 \label{eq:im-log-b-window}
\end{equation}
Choose \(|t|\) smaller than the separation from \(\sigma_n\) to its
neighbouring levels, and take \(s\in(\sigma_n,\sigma_n+|t|)\). The window in
\eqref{eq:im-log-b-window} contains only this level, giving
\(\operatorname{Im}\M(s+i0,t)=g\,e^{\operatorname{Re}\log B}\sin(\pi w_n)/(stu)\).
The forward discontinuity vanishes between the poles, so
\eqref{eq:nonforward-bound} requires \(\sin(\pi w_n)=0\). Thus every weight
is a positive integer. If
\(w_n\ge2\), \(B\) has a pole of order \(w_n\) at \(s=\sigma_n\). Its
coefficient contains the factor \((\sigma_n+u)^{w_n}=(-t)^{w_n}\) and is
nonzero for generic small \(t\neq0\). Then \(\operatorname{Im}\M(s+i0,t)\)
contains \(\delta^{(w_n-1)}(s-\sigma_n)\), which is not a measure, again
contradicting \eqref{eq:nonforward-bound}. Therefore
\begin{equation}
 w_n=1\quad\text{for all }n,\qquad
 B=\prod_n
 \frac{(\sigma_n+s)(\sigma_n+t)(\sigma_n+u)}
 {(\sigma_n-s)(\sigma_n-t)(\sigma_n-u)}.
 \label{eq:closed-B-unit-product}
\end{equation}
Simple poles therefore follow from (H1)--(H4). The argument only needs
nonzero momentum transfer arbitrarily close to the forward limit.

For illustration, in units \(M=1\), consider the family
\begin{equation}
 B_{\lambda,w}(s,t,u)=B_{\rm VS}(\lambda s,\lambda t,\lambda u)^{w}
 \label{eq:vs-power-deformation}
\end{equation}
For \(w>0\) and \(0<\lambda\le1\), it obeys the all-order ansatz, with
\(\alpha_{2k+1}=w\lambda^{2k+1}\,
2\zeta(2k+1)/(2k+1)\). Its logarithmic measure
\(\dd\nu(\sigma)=w\sum_{n\ge1}\delta(\sigma-n/\lambda)\dd\sigma\)
is positive, with support above the gap. The family also has Regge behaviour
\(|B_{\lambda,w}|\sim|s|^{2w\lambda t}\) at fixed \(t<0\).
Thus the ansatz and forward positivity allow every positive \(w\), whereas
non-forward positivity selects \(w=1\).

\paragraph{Fixing the spectrum.}
We now use (H5): only finitely many spins contribute at the first massive
pole. Its residue is then a polynomial in \(t\), with the positive
partial-wave expansion
\begin{equation}
 R_1(t)\equiv\operatorname*{Res}_{s=\sigma_1}
 \M_{\rm sub}(s,t,-s-t)
 =-\sum_{\ell=0}^{\ell_1}g_\ell^2
 P_\ell^{(D)}\left(1+\frac{2t}{\sigma_1}\right),
 \qquad g_\ell^2\ge0.
\end{equation}
For \(t>0\), the finite sum is strictly negative for a nonzero residue.
The polynomial therefore has no positive zeros. This is where finite spin
is needed: an infinite partial-wave expansion need not converge throughout
the positive \(t\) axis.

Taking the first residue of the product gives
\begin{equation}
 R_1(t)=-\frac{2g}{(\sigma_1-t)(2\sigma_1+t)}
 \prod_{n\ne1}
 \frac{(\sigma_n+\sigma_1)(\sigma_n+t)(\sigma_n-\sigma_1-t)}
 {(\sigma_n-\sigma_1)(\sigma_n-t)(\sigma_n+\sigma_1+t)}.
 \label{eq:first-pole-subtracted-residue}
\end{equation}
Polynomiality requires cancellation of the apparent pole at
\(t=\sigma_1\), so a level at \(2\sigma_1\) must be present. A level between
\(\sigma_1\) and \(2\sigma_1\) would instead produce an uncancelled zero
in \(0<t<\sigma_1\), contradicting positivity. Hence
\(\sigma_2=2\sigma_1\). More generally, every apparent positive zero at
\(t=\sigma_n-\sigma_1\) must coincide with a positive pole. Therefore
\begin{equation}
 \sigma_n-\sigma_1\in\{\sigma_m\}_{m\ge1}.
 \label{eq:sigma-shift-condition}
\end{equation}
This fixes the spectrum inductively. If the first \(n-1\) levels are
\(j\sigma_1\), then \(\sigma_n-\sigma_1\) is one of them and is larger
than \((n-2)\sigma_1\); it must equal \((n-1)\sigma_1\).
The sequence cannot terminate, since the last positive pole of the residue
would remain uncancelled. Thus
\begin{equation}
 \sigma_n=n\sigma_1,\qquad n=1,2,\ldots,
 \label{eq:equally-spaced-spectrum}
\end{equation}
Substitution into the moment formula gives
\begin{equation}
 \alpha_{2k+1}=\frac{2\zeta(2k+1)}{2k+1}
 \frac{1}{\sigma_1^{2k+1}}.
 \label{eq:alpha-equally-spaced-spectrum}
\end{equation}
These coefficients identify the Virasoro-Shapiro analytic germ and its
meromorphic continuation, with squared-mass spacing \(\sigma_1\).
Since \(\sigma_1\ge M^2\),
\(c_{0,0}M^6/(8\pi G)=2\zeta(3)(M^2/\sigma_1)^3\le2\zeta(3)\), with
equality when \(\sigma_1=M^2\).

\paragraph{Relation to the numerical bounds.}
NL3 and NL6 impose finite subsets of the full nonlinear relations. They
correlate non-forward Wilson coefficients with the forward moments, so
positivity turns these correlations into bounds on the low-energy data.
With the dispersive input fixed, adding more relations can only reduce the allowed
set. The spectral assumptions (H4) and (H5) enter only the analytic
derivation of the string spectrum above.

\section{Discussion}
\label{sec:discussion}

In this work we studied the Virasoro-Shapiro amplitude using dispersive
bootstrap methods. With the gravity pole kept explicitly, fixed-\(t\)
dispersion relations, crossing null constraints, and partial-wave positivity
give bounds on the leading low-energy coefficients that agree at our
reference truncation with ref.~\cite{Albert:2024yap}. We then imposed Virasoro-inspired nonlinear
relations among Wilson coefficients and found that they shrink the allowed
region toward the Virasoro-Shapiro trajectory. In the gravity-pole-subtracted
setup at \(c_{0,0}=2\zeta(3)\), the same relations reduce the allowed region
to a small island containing the Virasoro-Shapiro point. Imposed to all
orders, the ansatz and partial-wave positivity at nonzero momentum transfer
force unit weights, and with isolated exchanges and finitely many spins at the
first massive level they select the string spectrum. These two assumptions are
not imposed numerically, so the analytic result does not imply that the
numerical island shrinks to a point.

Our analysis is restricted to the tree-level bootstrap. In this regime
we use positivity of the absorptive part, or equivalently positive
partial-wave spectral densities. A more complete nonperturbative bootstrap can
instead impose full unitarity, including both absorptive and non-absorptive
parts of the amplitude, as in ref.~\cite{Guerrieri:2021ivu}. In such a setup,
there may still remain a small gap between the bootstrap-allowed region and the
exact string amplitude. It would therefore be interesting to ask whether
additional stringy input can close this gap and further isolate the solution.

It would also be interesting to extend the present analysis beyond flat
space. Recent work has proposed monodromy relations for string amplitudes in
AdS, where the flat-space Veneziano monodromy relations are recovered in the
small-curvature limit \cite{Alday:2025cxr}. This raises the natural question
of whether analogous stringy constraints, together with CFT crossing and
unitarity, can uniquely determine AdS string amplitudes. Developing such a
bootstrap would provide a bridge between the flat-space S-matrix approach
studied here and the AdS/CFT bootstrap of stringy correlators.

\paragraph{Note added.}
Reference~\cite{Berman:2026gravityparity} derived related nonlinear
constraints in four-dimensional maximally supersymmetric gravity from
higher-point factorization and a parity condition. Its sections~6.3
and~6.4 show, numerically and analytically, that requiring finitely many
spins at the first mass level leaves only the Virasoro-Shapiro amplitude,
with equally spaced squared masses; this parallels the role of (H5) in
section~\ref{subsec:stripped-analytic-structure}.
Reference~\cite{Huang:2026infrared} has since established the corresponding
exponential forms to all orders in this setting, using six-point factorization, maximal
supersymmetry, scalar parity and a discrete R-symmetry. Its gravitational
analysis likewise identifies finite spin at the first massive pole as an
additional input selecting the Virasoro-Shapiro spectrum. These results
provide further motivation for the ansatz studied here.

\acknowledgments
I am grateful to Longqi Shao for carefully reading the draft and to Shilin Wan
for useful discussions on the analytic bootstrap. I also thank Anna Tokareva
and Alexey Koshelev for useful discussions and comments.

\appendix

\section{Numerical details}
\label{app:numerical-details}

The numerical problem is to find a positive spectral density compatible
with the chosen sum rules and nonlinear relations. We approximate it by
nonnegative weights on a finite \((\mu,\ell)\) grid. Once the required lower
Wilson coefficients are fixed, each optimization is an LP,
\begin{equation}
 \min_{\rho\ge0}f^T\rho,\qquad E\rho=b,
 \label{eq:revision-primal-lp}
\end{equation}
with dual \(\max b^Ty\) subject to \(E^Ty\le f\). The matrix \(E\) contains
the moment and crossing-null kernels, the sampled gravity-pole kernels when
present, and the nonlinear relations made linear by the fixed coefficients.
The grid approximates support on \(M^2\le\mu<\infty\); its largest node is a
numerical choice, not an upper mass assumption. The code and data are
available at
\begin{center}
  \url{https://github.com/yongjunx23-del/A-Dispersive-Bootstrap-for-the-Virasoro-Shapiro-Amplitude}.
\end{center}
The folder \texttt{revision\_v2} contains the numerical companion of this
version, with the scan configurations, audited records and figure scripts;
the folder \texttt{v1} contains the implementation used for the first version.

\paragraph{Solving and checking the LPs.}
We use SDPX \cite{SDPX:2026}\footnote{The first version used Gurobi with double-precision arithmetic (Float64).}
(\url{https://github.com/yongjunx23-del/SDPX.rs}) with the precisions and
thresholds in table~\ref{tab:solver-settings}. Kernels are evaluated at
512 bits, the LPs are solved at 256 bits, and the final results are audited
independently at 768 bits. Solver accuracy and boundary-scan resolution
are separate controls.

\begin{table}[htbp]
 \centering
 \small
 \begin{tabular}{lcc}
 \toprule
 Setting & With the graviton pole & Pole-subtracted \\
 \midrule
 Kernel precision & 512 bits & 512 bits \\
 LP precision & 256 bits & 256 bits \\
 Independent audit precision & 768 bits & 768 bits \\
 Solver feasibility threshold & \(10^{-36}\) & \(10^{-30}\) \\
 Solver absolute and relative gap thresholds & \(10^{-36}\) & \(10^{-30}\) \\
 Original-equation and full-grid audit threshold & \(10^{-25}\) & \(10^{-25}\) \\
 \bottomrule
 \end{tabular}
 \caption{Arithmetic precision and acceptance thresholds. The independent
 audit evaluates the original equations and all columns of the declared
 spectral grid. Boundary-scan resolution is a separate parameter.}
 \label{tab:solver-settings}
\end{table}

To avoid placing every spectral column in the initial LP, we begin with a
restricted set. After solving it, we evaluate the dual slack \(f-E^Ty\) on
the full energy--spin grid, add the most violated columns and solve again.
This is repeated until no column is violated. The independent audit then
checks the original equations and every declared column. A finite optimum
requires nonnegative weights, normalized equality residuals and normalized
objective gap below \(10^{-25}\), and nonnegative dual slack. Exclusion requires a
dual ray with \(b^Ty>0\) and \(E^Ty\le0\) on every column. For these
finite-bound and feasibility scans, inconclusive audit checks are recorded
as unresolved and do not define the plotted bounds.
In the spectrum plots, weights below \(10^{-22}\rho_{\max}\) are treated as
zero. They lie at the interior-point numerical floor,
\(10^{-24}\)--\(10^{-35}\rho_{\max}\), separated from the displayed support.

SDPX accepts an LP directly, whereas SDPB
\cite{Simmons-Duffin:2015qma,Landry:2019qug} uses a polynomial matrix
representation. Table~\ref{tab:sdpb-sdpx} compares the solvers on two LPs,
using identical inputs for each test, one core, 256 bits and common
\(10^{-30}\) tolerances. Their objective values
agree within \(3\times10^{-30}\); the median SDPX time is 21--47 times
shorter, including conversion in the SDPB time. This comparison concerns
only these two LPs.

\begin{table}[htbp]
 \centering
 \small
 \begin{tabular}{ccccc}
 \toprule
 columns$\times$rows & SDPX s (it.) & SDPB total s (it.) & ratio & $|\Delta\,\mathrm{obj}|$\\
 \midrule
 $816\times29$ & 2.3 (39) & 105.2 (190) & 46.5 & $1.2\times10^{-30}$\\
 $3232\times44$ & 28.9 (68) & 606.7 (232) & 21.0 & $2.8\times10^{-30}$\\
 \bottomrule
 \end{tabular}
 \caption{SDPX and SDPB on two pole-subtracted NL6 string-slice LPs, with identical inputs for each solver comparison, 256-bit arithmetic and $10^{-30}$ thresholds, using 1 solver core. Times and iteration counts are medians of 3 sequential repeats; SDPB total time includes pmp2sdp conversion. The last column gives the largest objective difference across the repeats.}
 \label{tab:sdpb-sdpx}
\end{table}

\paragraph{Gravitational scans.}
For the dispersive region we maximize the support function in 192
directions, \(\theta_j=2\pi j/192\), and take the convex hull of the audited
optima. The 48- and 96-direction meshes are nested subsets of this mesh.

For NL3 we fix \(\gamma=c_{0,0}M^6/(8\pi G)\) and minimize and maximize
\(c_{1,0}\). At fixed \(\gamma\),
\eqref{eq:ansatz-NL1}--\eqref{eq:ansatz-NL3} are linear in the spectral
weights. In addition to the original slices and their midpoints, we sample
\(\gamma=2\zeta(3)\alpha^3\) at
\(\alpha=0.10,0.15,\ldots,0.90\) and
\(0.925,0.95,0.97,0.98,0.99,1\). Near the string endpoint we add
\(\gamma=2\zeta(3)+k\,5\times10^{-5}\), \(k=-10,\ldots,7\), and further
slices toward the end of NL3. Bisection between the last nonempty and first
empty slices determines the endpoint bracket.

For NL6 we also fix \(c_{1,0}\), making
\eqref{eq:ansatz-NL1}--\eqref{eq:ansatz-NL6} linear. At each nonempty NL3
slice \([a,b]\), the initial trials are
\(c_{1,0}=a+(k+\tfrac12)(b-a)/17\), \(k=0,\ldots,16\), the two
endpoints, and the string-reference value
\(-2\zeta(5)[\gamma/(2\zeta(3))]^{5/3}\). Previously audited trials on
the same slice are retained. We bisect each adjacent feasible/excluded
pair, audit the midpoint, and replace the endpoint with the same
classification. Refinement stops at a bracket width of \(0.5\%\) of the
explored feasible span or after 24 steps. An unresolved midpoint stops that
bracket and is never counted as excluded. Separate detected windows are
kept. Near the observed tip, eight further normalization slices are tested
between the last nonempty slice and the next tested normalization, with
21 coefficient trials around the extrapolated window at each slice.

\paragraph{Pole-subtracted scans.}
At \(c_{0,0}=2\zeta(3)\), in units \(8\pi G=M=1\), the NL0 and NL3
regions come from support-function LPs in the \((c_{1,0},c_{2,0})\) plane,
using 192 and 384 directions, respectively. For NL6 we instead fix
\(x=M^4c_{1,0}/c_{0,0}=x_{\rm VS}+i\,2\times10^{-5}\), where
\(x_{\rm VS}=-\zeta(5)/\zeta(3)\), and minimize and maximize \(c_{2,0}\).
We extend the range of \(i\) until the end slices are empty and bisect both
edges five times, giving a resolution of \(6\times10^{-7}\) in \(x\).
Each checked nonempty interval is refined fourfold for the main figure,
to a spacing of at most \(5\times10^{-6}\).

Higher-NL boundaries require many LPs because the initial scan must be
followed by repeated boundary tests. A bracket of width \(\Delta\) takes
\(\lceil\log_2(\Delta/\epsilon)\rceil\) bisections to reach \(\epsilon\).
This resolves a detected transition, while a finite initial scan may miss
other disconnected windows.

\subsection{Convergence checks}
\label{app:revision-convergence}

\begin{figure}[htbp]
 \centering
 \includegraphics[width=0.96\textwidth]{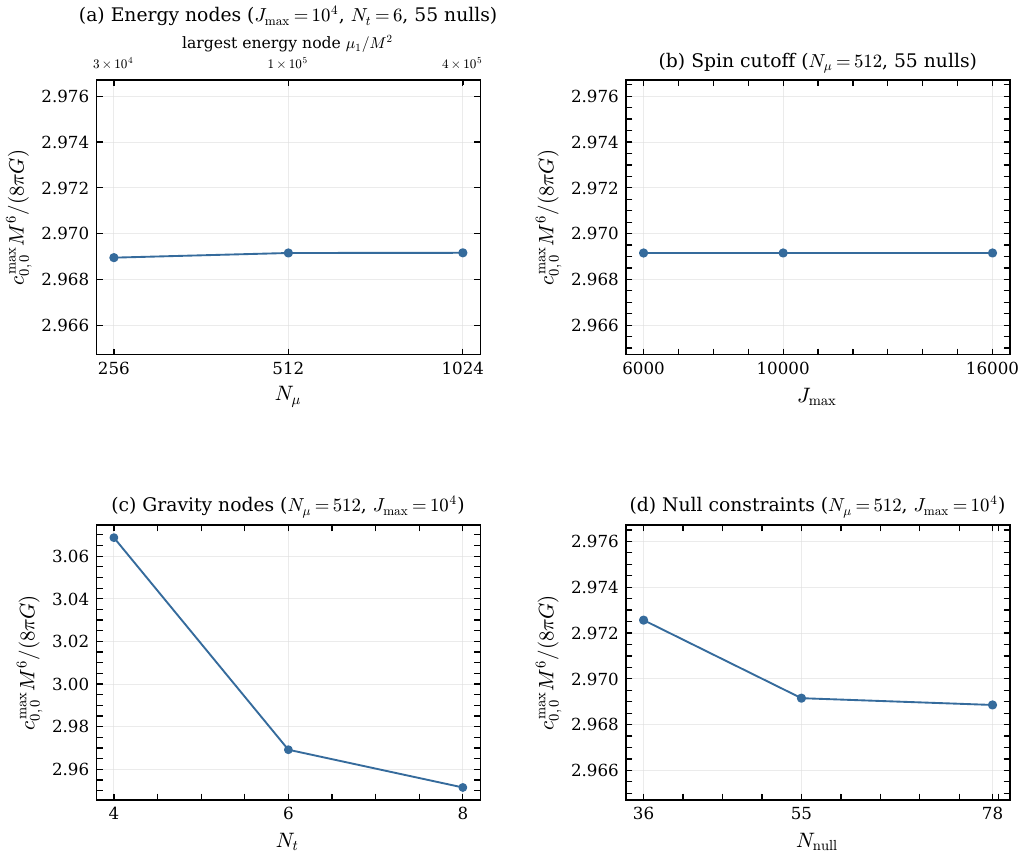}
 \caption{Linear maximum of $c_{0,0}M^6/(8\pi G)$, changing one truncation at a
 time about $(N_\mu,J_{\max},N_t,N_{\rm null})=(512,10^4,6,55)$: (a) energy
 counts $N_\mu=256,512,1024$, with the largest numerical energy node
 $\mu_1=\csc^2[\pi/(2N_\mu)]M^2$ on the top
 axis; (b) spin cutoff; (c) gravity nodes; (d) null constraints. Panels (a),
 (b), and (d) share one vertical window.}
 \label{fig:gravity-convergence-revision}
\end{figure}

\begin{figure}[htbp]
 \centering
 \includegraphics[width=0.96\textwidth]{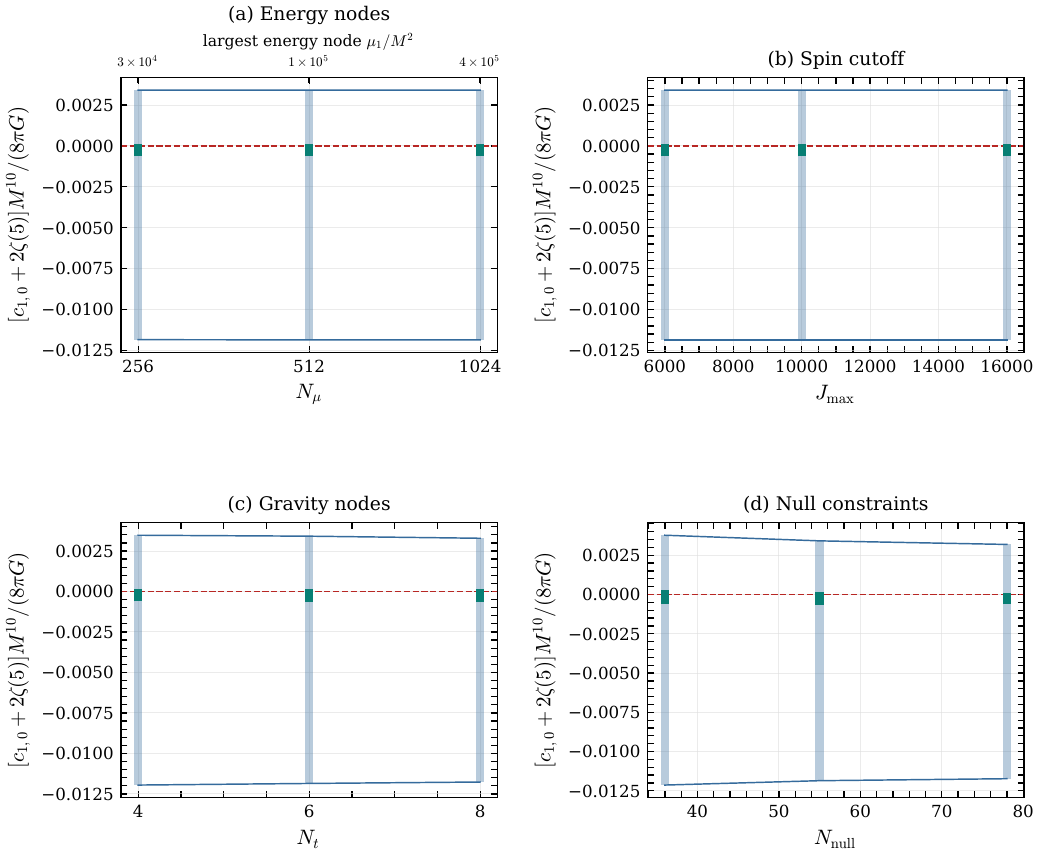}
 \caption{Nonlinear gravitational results on the slice
 $c_{0,0}M^6/(8\pi G)=2\zeta(3)$, varying one setting about
 $(N_\mu,J_{\max},N_t,N_{\rm null})=(512,10^4,6,55)$.
 Light bars: NL3 interval of $c_{1,0}+2\zeta(5)$; dark bars:
 span of feasible NL6 tests. The top axis of panel (a) gives the highest
 grid energy. The dashed line is the string value.}
 \label{fig:gravity-nl-convergence}
\end{figure}

\begin{figure}[htbp]
 \centering
 \includegraphics[width=0.96\textwidth]{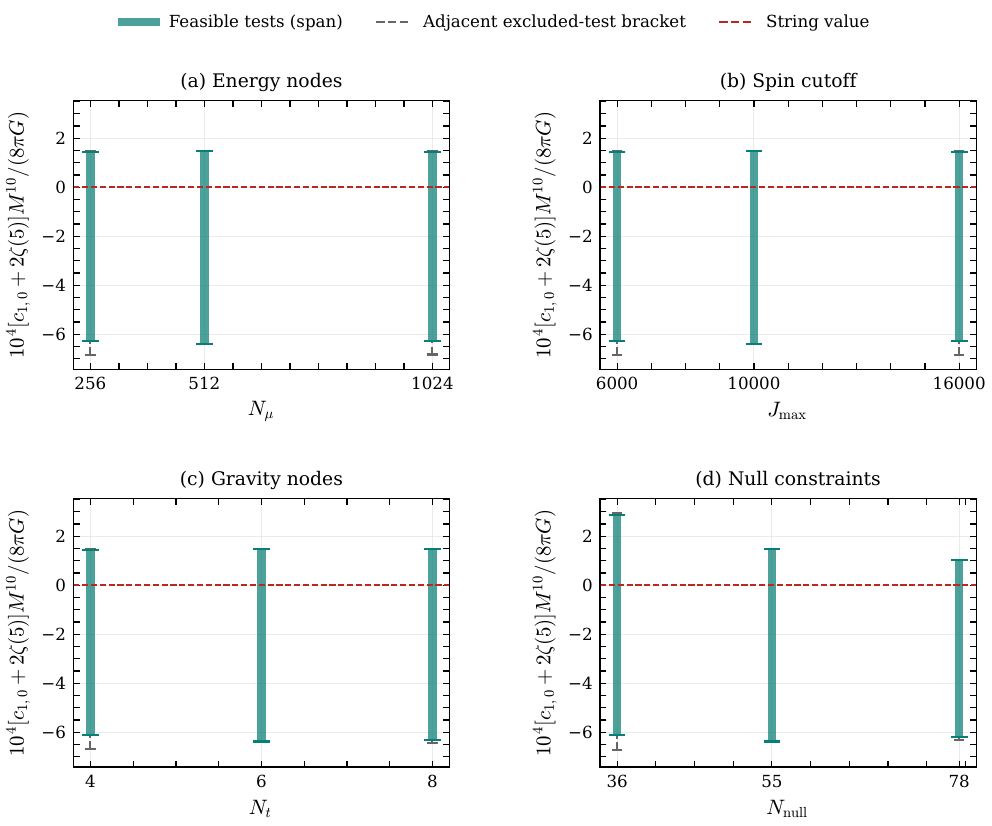}
 \caption{The gravitational NL6 tests of
 figure~\ref{fig:gravity-nl-convergence} on a common expanded scale.
 Solid bars span the feasible tests; grey dashed extensions reach the
 nearest excluded tests on either side. These extensions quantify the
 scan resolution and do not certify a continuous NL6 boundary.
 The red dashed line is the string value.}
 \label{fig:gravity-nl6-convergence}
\end{figure}

\begin{figure}[htbp]
 \centering
 \includegraphics[width=0.96\textwidth]{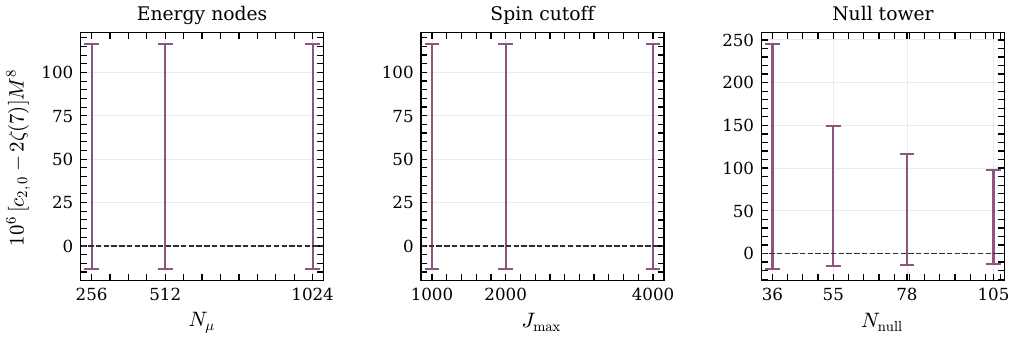}
 \caption{Pole-subtracted NL6 interval of $c_{2,0}$ at the string values of
 $c_{0,0}$ and $c_{1,0}$, shown as $10^6[c_{2,0}-2\zeta(7)]$ ($M=1$), changing
 one setting about $(N_\mu,J_{\max},N_{\rm null})=(512,2000,78)$.}
 \label{fig:pole-convergence-revision}
\end{figure}

\begin{figure}[htbp]
 \centering
 \includegraphics[width=0.96\textwidth]{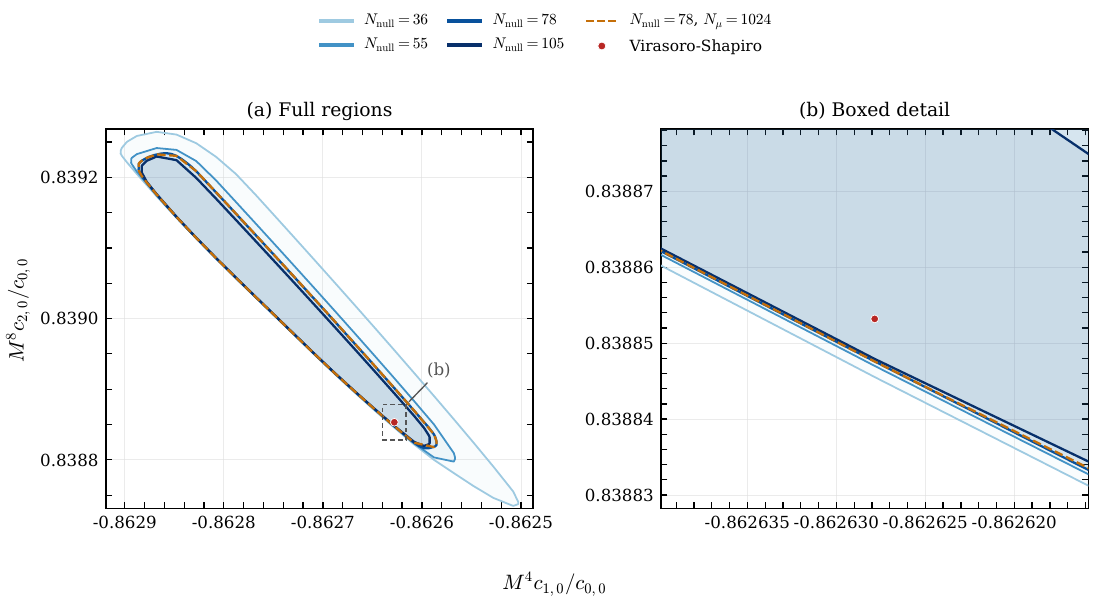}
 \caption{Pole-subtracted NL6 islands at $c_{0,0}=2\zeta(3)$ for four null
 towers and a finer energy grid. Both panels show the coefficient ratios
 $M^4c_{1,0}/c_{0,0}$ and $M^8c_{2,0}/c_{0,0}$ directly. Panel (a) shows
 the full displayed islands; its dashed box marks the extent of panel (b).
 The islands shrink as null constraints are added, and the red point marks
 the string reference. Outlines close at the extreme checked slices.}
 \label{fig:pole-band-truncations}
\end{figure}

We vary the energy resolution, spin cutoff, gravity samples and null tower
in figures~\ref{fig:gravity-convergence-revision}--\ref{fig:pole-band-truncations}.
Adding constraints to the same grid can only shrink the feasible set;
adding spectral states with the constraints fixed can only enlarge it.
The energy grids are nested when \(N_\mu\) is doubled; the gravity node
sets for different \(N_t\) are not.
The null tower contains \((k,n)=(2m,n)\), with \(1\le n\le p\) and
\(0\le m<n\): \(p(p+1)/2\) rows, or 36, 55, 78 and 105 for
\(p=8,10,12,14\). Adding null constraints strengthens the bounds. We do
not push the null tower to its optimum: higher null constraints probe
higher spins, so each enlargement also requires a larger spin cutoff.
Only audited outcomes enter the plots; the companion
contains the full tables and records.

\paragraph{Sampling the graviton pole.}
Larger \(N_t\) can strengthen a finite bound, but also probes nearer the
forward pole: the closest Chebyshev node has
\(|t_1|=\sin^2[\pi/(4N_t)]\). Resolving these conditions requires an
adequately dense energy grid, support at higher energies and a larger
\(J_{\max}\), since the pole is built from arbitrarily high energy and
spin \cite{Haring:2024wyz}. Thus the energy support and spin cutoff must
be increased together with \(N_t\). Smeared calculations likewise need
greater mass and spin resolution as their functional basis is enlarged
\cite[section~3.2 and appendix~A]{Caron-Huot:2021rmr}.
Here we compare \(N_t=4,6,8\), whose node sets are not nested, to assess
sensitivity at the stated settings; we do not seek the optimal
gravitational bound.

\paragraph{Nonlinear gravitational regions.}
Figures~\ref{fig:gravity-nl-convergence} and
\ref{fig:gravity-nl6-convergence} show the NL3/NL6 ranges and the resolved
NL6 scale. Dashed extensions retain the adjacent feasible/excluded brackets.

On this slice, changing the energy count to \(N_\mu=256\) or 1024, or the
spin cutoff to \(J_{\max}=6000\) or 16000, changes the width of the
feasible NL6 window by at most \(1.7\%\), and \(N_t=4\) and 8 change it by
\(3.8\%\) and \(0.6\%\). Increasing the null tower from 55 to 78 rows
narrows it by \(8.1\%\). Removing the sampled pole rows altogether widens
the NL3 interval by \(7.7\%\) but changes the NL6 window by only
\(0.6\%\), so at this truncation the NL6 window is set mainly by the
nonlinear relations and the moment and null rows.

\paragraph{Pole-subtracted island.}
At the string values of \(c_{0,0}\) and \(c_{1,0}\), doubling the energy count from 512 to 1024 changes the NL6 width by \(0.0923\%\), and doubling the spin cutoff from 2000 to 4000 changes it by \(0.00257\%\). Increasing the null tower from 78 to 105 rows reduces the width by \(14.5\%\). Thus energy and spin are locally stable on these tests, while the null tower still narrows the island.

\FloatBarrier

\section{A fixed-\texorpdfstring{\(a\)}{a} derivation of null constraints}
\label{app:fixed-a-null}

In this appendix we describe a convenient way to generate the locality, or
crossing-null, constraints in the massless case. The construction is based on
the fixed-\(a\) form of the crossing-symmetric dispersion relation
\cite{Auberson:1972prg,Mahoux:1974ej,Sinha:2020win}; see
section~\ref{subsec:fixed-t} for further references. We use
the crossing-symmetric variables
\[
  x=st+su+tu,
  \qquad
  y=stu,
  \qquad
  a=\frac{y}{x},
  \qquad
  s+t+u=0 .
\]
At fixed \(a\), locality of the low-energy expansion requires the coefficient
of \(x^m\) to be a polynomial in \(a\) of degree at most \(m\). Therefore, any
terms on the dispersive side proportional to higher powers of \(a\) must
vanish. These vanishing conditions give the crossing-null constraints.

For even \(k\ge0\), write \(m=k/2\). The low-energy side of the fixed-\(a\)
dispersion relation is
\begin{equation}
  -c^{\rm low}_{2m}(a)
  =
  \frac{(-1)^m}{m!}
  \partial_x^m \M(x,a)\big|_{x=0}.
  \label{eq:fixed-a-low-side}
\end{equation}
Using the local expansion
\begin{equation}
  \M(x,a)
  =
  \frac{1}{ax}
  +
  \sum_{r=0}^{\infty}\sum_{n=0}^{r}
  c_{r,n}x^r a^n ,
\end{equation}
and treating the gravity pole separately, we obtain
\begin{equation}
  -c^{\rm low}_{2m}(a)
  =
  (-1)^m
  \sum_{n=0}^{m}c_{m,n}a^n .
  \label{eq:fixed-a-low-polynomial}
\end{equation}
Thus the low-energy side is a polynomial in \(a\) of degree at most \(m\).

The high-energy side can be written as
\begin{equation}
  c^{\rm high}_{2m}(a)
  =
  \left\langle
  \frac{
  P_\ell^{(D)}\!\left(\sqrt{\frac{\mu+3a}{\mu-a}}\right)}
  {\mu^{3m}}
  (\mu-a)^{m-1}
  (2\mu-3a)
  \right\rangle .
  \label{eq:fixed-a-high-side}
\end{equation}
Introducing
\begin{equation}
  z=\frac{a}{\mu},
\end{equation}
the kernel becomes
\begin{equation}
  c^{\rm high}_{2m}(a)
  =
  \left\langle
  \frac{1}{\mu^{2m}}\,
  F_{m,J}^{(D)}(z)
  \right\rangle ,
  \qquad
  F_{m,J}^{(D)}(z)
  =
  (2-3z)(1-z)^{m-1}
  P_J^{(D)}\!\left(
  \sqrt{\frac{1+3z}{1-z}}
  \right).
  \label{eq:fixed-a-F-kernel}
\end{equation}
We expand
\begin{equation}
  F_{m,J}^{(D)}(z)
  =
  \sum_{n=0}^{\infty}
  Q_{m,n}^{(J,D)} z^n,
  \qquad
  Q_{m,n}^{(J,D)}
  =
  [z^n]F_{m,J}^{(D)}(z).
  \label{eq:fixed-a-Qmn-definition}
\end{equation}
Then
\begin{equation}
  c^{\rm high}_{2m}(a)
  =
  \sum_{n=0}^{\infty}
  a^n
  \left\langle
  \frac{Q_{m,n}^{(J,D)}}{\mu^{2m+n}}
  \right\rangle .
  \label{eq:fixed-a-high-taylor}
\end{equation}
Since the fixed-\(a\) dispersion relation gives
\begin{equation}
  -c^{\rm low}_{2m}(a)=c^{\rm high}_{2m}(a),
\end{equation}
comparison with \eqref{eq:fixed-a-low-polynomial} gives the Wilson-coefficient
sum rules
\begin{equation}
  \left\langle
  \frac{Q_{m,n}^{(J,D)}}{\mu^{2m+n}}
  \right\rangle
  =
  (-1)^m c_{m,n},
  \qquad
  0\le n\le m ,
  \label{eq:fixed-a-wilson-sum-rule}
\end{equation}
and the null constraints
\begin{equation}
 \left\langle
  \frac{Q_{m,n}^{(J,D)}}{\mu^{2m+n}}
  \right\rangle
  =
  0,
  \qquad
  n>m .
  \label{eq:fixed-a-null-constraints}
\end{equation}

\section{Locality and power sums}
\label{app:power-sums-locality}

In this appendix we collect two simple consequences of the power-sum
identities that are useful for the Virasoro-inspired ansatz. Define
\begin{equation}
  p_n=s^n+t^n+u^n.
\end{equation}
Newton's identities give
\begin{equation}
  p_n=-x\,p_{n-2}+y\,p_{n-3}.
  \label{eq:appendix-power-sum-recursion}
\end{equation}

First, within the assumed separable logarithm, we show why even power sums
are absent from \(\log B\). For every
\(k\ge1\), the even power sums have the structure
\begin{equation}
  p_{2k}
  =
  2(-1)^k x^k
  +y^2 G_k(x,y^2),
  \label{eq:appendix-even-power-structure}
\end{equation}
where \(G_k\) is a polynomial, with the second term absent for \(k=1,2\). For
example,
\begin{equation}
  p_2=-2x,
  \qquad
  p_4=2x^2,
  \qquad
  p_6=-2x^3+3y^2 .
\end{equation}
The proof is by induction. If \eqref{eq:appendix-even-power-structure} holds
for \(p_{2k}\), then
\begin{equation}
  p_{2k+2}
  =
  -x\,p_{2k}
  +y\,p_{2k-1}.
\end{equation}
Odd power sums are divisible by \(y\), since \(p_1=0\), \(p_3=3y\), and the
recursion preserves divisibility by \(y\) for odd \(n\). Hence
\(y\,p_{2k-1}\) is divisible by \(y^2\), while the first term gives
\(2(-1)^{k+1}x^{k+1}\). This proves
\eqref{eq:appendix-even-power-structure}.

Now suppose that an even power sum \(\beta_{2k}p_{2k}\) appears in \(\log B\).
Already at linear order in \(\beta_{2k}\),
\begin{equation}
  B=1+\beta_{2k}p_{2k}+\cdots .
\end{equation}
Dividing by \(stu=y=xa\), this gives
\begin{equation}
  \frac{\beta_{2k}p_{2k}}{y}
  =
  2(-1)^k\beta_{2k}\frac{x^k}{y}
  +\text{local terms}
  =
  2(-1)^k\beta_{2k}\frac{x^{k-1}}{a}
  +\text{local terms}.
\end{equation}
The term \(x^{k-1}/a=x^k/y\) is not a polynomial in \(s,t,u\). It is an
additional low-energy singularity beyond the isolated gravity pole. Therefore
locality of the pole-subtracted EFT requires all even-power coefficients in
this separable power-sum ansatz to vanish. Local products such as \(y^2\) in
a more general \(\log B\) remain allowed; their exclusion is part of the
additional ansatz, rather than a consequence of locality.

Second, we derive the all-order linear relation between the odd coefficients
\(\alpha_{2k+1}\) and the Wilson coefficients \(c_{m,0}\). The odd power sums
have the leading structure
\begin{equation}
  p_{2k+1}
  =
  (2k+1)(-1)^{k+1}x^{k-1}y
  +O(y^2),
  \qquad k\ge1 .
  \label{eq:appendix-odd-leading-structure}
\end{equation}
Since \(y=xa\), this is
\begin{equation}
  p_{2k+1}
  =
  (2k+1)(-1)^{k+1}x^k a
  +O(a^2).
\end{equation}
Substituting into
\begin{equation}
  \log B
  =
  \sum_{k\ge1}\alpha_{2k+1}p_{2k+1},
\end{equation}
and using \(\M=B/(xa)\), we find
\begin{equation}
  c_{k-1,0}
  =
  (2k+1)(-1)^{k+1}\alpha_{2k+1},
  \qquad k\ge1 .
  \label{eq:appendix-linear-alpha-relation-k}
\end{equation}
Equivalently,
\begin{equation}
c_{m,0}
  =
  (2m+3)(-1)^m\alpha_{2m+3},
  \qquad m\ge0 .
  \label{eq:appendix-linear-alpha-relation}
\end{equation}
The first few cases are
\begin{equation}
  c_{0,0}=3\alpha_3,
  \qquad
  c_{1,0}=-5\alpha_5,
  \qquad
  c_{2,0}=7\alpha_7,
  \qquad
  c_{3,0}=-9\alpha_9 .
\end{equation}

\section{Regge behaviour within the all-order ansatz}
\label{app:positivity-regge}

We use the representation \eqref{eq:closed-logB-moment-form}, which follows
from (H1) and (H2), with
\(L_\sigma(z)=\log[(\sigma+z)/(\sigma-z)]\). Since \(L_\sigma\) is odd and
\(u=-s-t\),
\begin{equation}
 \log B(s,t)=\int d\nu(\sigma)\,
 \big[L_\sigma(s)+L_\sigma(t)-L_\sigma(s+t)\big].
 \label{eq:app-logb-st}
\end{equation}
Expanding to first order in \(t\) and comparing with the forward limit of
\(\M_{\rm sub}=g(B-1)/(stu)\) gives the exact identity
\begin{equation}
 \partial_t\log B(s,t)\big|_{t=0}
 =\int d\nu(\sigma)\left[\frac2\sigma-\frac{2\sigma}{\sigma^2-s^2}\right]
 =-\frac{s^2}{g}\,\M_{\rm sub}(s,0),
 \qquad
 \frac{\M_{\rm sub}(s,0)}{g}=\int\frac{2\,d\nu(\sigma)}{\sigma(\sigma^2-s^2)} .
 \label{eq:app-forward-derivative}
\end{equation}
For fixed \(|t|<\sigma_1\) and \(|s|\to\infty\) in a sector
\(\epsilon\le\arg s\le\pi-\epsilon\), the second-order remainder of
\(L_\sigma(s)-L_\sigma(s+t)\) is bounded by
\(t^2\int d\nu\,\sigma|s|/|\sigma^2-s^2|^2\). This is smaller than the
first-order term by a factor \(O(|t/s|)\), and
\(L_\sigma(t)-tL'_\sigma(0)=O(|t|^3/\sigma^3)\) gives a bounded contribution.
Therefore
\begin{equation}
 \log B(s,t)=-\frac{t\,s^2}{g}\,\M_{\rm sub}(s,0)\,\big[1+O(|t/s|)\big]+O(1).
 \label{eq:app-regge-forward}
\end{equation}
Within the ansatz, the fixed-\(t\) high-energy behaviour is the exponential of
the forward amplitude. For a tail
\(d\nu=(\kappa+O(\sigma^{-\delta}))\,d\sigma\), with \(\delta>0\),
splitting the integral in
\eqref{eq:app-forward-derivative} at \(\sigma=|s|\) gives
\(s^2\M_{\rm sub}(s,0)/g=-2\kappa\log(|s|/M^2)+O(1)\).
The bounded remainder in this estimate uses the stated constant-tail
asymptotics.

For a power-law tail \(d\nu=\kappa\sigma^\beta d\sigma\), \(0<\beta<2\), the
Mellin integral
\(\int_0^\infty\sigma^{\beta-1}(\sigma^2+a^2)^{-1}d\sigma
=\tfrac{\pi}{2}a^{\beta-2}/\sin(\pi\beta/2)\) with \(a=-is\) gives
\begin{equation}
 \frac{s^2}{g}\M_{\rm sub}(s,0)=-\frac{\pi\kappa(-is)^\beta}{\sin(\pi\beta/2)}+O(1),
 \qquad
 \log B\simeq\frac{\pi\kappa\,t\,(-is)^\beta}{\sin(\pi\beta/2)} .
 \label{eq:app-power-tail}
\end{equation}
Let \(\varphi=\arg s-\pi/2\in(-\pi/2,\pi/2)\). Then
\(\operatorname{Re}\log B\simeq\pi\kappa t|s|^\beta\cos(\beta\varphi)/\sin(\pi\beta/2)\).
For \(t<0\) and \(\beta\le1\), this is nonpositive throughout the upper
half-plane, so \(B\) is bounded and \(\M=gB/(stu)\) obeys the fixed-\(t\)
Regge bound used in the dispersion relations. For \(1<\beta<2\),
\(\cos(\beta\varphi)<0\) when \(|\varphi|>\pi/(2\beta)\), and \(|B|\) grows
like \(\exp(c|t||s|^\beta)\) near the real axis, violating that bound. The
logarithmic case \(\beta=0\) gives
\(s^2\M_{\rm sub}(s,0)/g\simeq-2\kappa\log(-is)\) and
\(|B|\sim|s|^{2\kappa t}\). For the Virasoro-Shapiro measure
\(\sum_n\delta(\sigma-n)\), \(s^2\M_{\rm sub}(s,0)/g=-\psi(1-s)-\psi(1+s)-2\gamma_E\),
which reproduces \(|B_{\rm VS}|\sim|s|^{2t}\).

These statements constrain only the large-\(\sigma\) behaviour of \(\nu\).
Adding or moving finitely many levels changes each \(\alpha_{2k+1}\) by a
finite amount but changes \(\log B\) at fixed \(t\) only by a bounded amount.
Hence, within the ansatz, Regge boundedness alone does not determine the
low-energy coefficients.
\bibliographystyle{JHEP}
\bibliography{biblio}
\end{document}